\documentclass[a4paper,12pt]{article}

\usepackage{a4}
\usepackage{graphics}
\usepackage{epsfig}
\usepackage{amssymb}
\usepackage{color}
\usepackage{cite}
\usepackage{epsf}
\usepackage{axodraw}
\usepackage{wasysym}

\definecolor{green3}{rgb}{0.,0.7,0.0}
\definecolor{red1}{rgb}{0.9,0,0}

\def\lsim{\raise0.3ex\hbox{$\;<$\kern-0.75em\raise-1.1ex\hbox{$\sim\;$}}}
\def\gsim{\raise0.3ex\hbox{$\;>$\kern-0.75em\raise-1.1ex\hbox{$\sim\;$}}}
\DeclareMathAlphabet{\scr}{U}{rsfs}{m}{n}

\begin{document}
\hspace*{\fill} 
MAN/HEP/2010/3\\
\vspace{1.5cm}
\begin{center}
{\Large\bf 
 CP violation in  sbottom decays 
}                       
\vspace{1.5cm}

{\sc Frank~F.~Deppisch\footnote{Email: frank.deppisch@manchester.ac.uk}}\\
\vspace*{.2cm} 
{\small \it School of Physics and Astronomy, University of Manchester, \\
Manchester M13 9PL, United Kingdom}
\vspace*{1.0cm}
            
{\sc Olaf~Kittel\footnote{Email: kittel@th.physik.uni-bonn.de}}\\
\vspace*{.2cm} 
{\small \it 
Departamento de F\'isica Te\'orica y del Cosmos and CAFPE, \\
Universidad de Granada, E-18071 Granada, Spain
}
\vspace*{.5cm}
\end{center}

\begin{abstract}\noindent
We study CP asymmetries in two-body decays of bottom squarks into charginos and 
tops. These asymmetries probe the SUSY CP phases of the sbottom and 
the chargino sector in the Minimal Supersymmetric Standard Model. 
We identify the MSSM parameter space where the CP asymmetries are sizeable,
and analyze the feasibility of their observation at the LHC.
As a result, potentially detectable CP asymmetries in sbottom decays 
are found, which motivates further detailed experimental studies for probing
the  SUSY CP phases.
\end{abstract}
\newpage

\section{Introduction}
\label{sec:Introduction}

Supersymmetry~(SUSY)~\cite{mssm} is a  well motivated theory to extend the Standard 
Model~(SM) of particle physics. SUSY models are not only favored by gauge coupling 
unification and naturalness  considerations, but are also attractive from the 
cosmological point of view. For instance the lightest SUSY particle~(LSP) is a good dark
 matter candidate if it is stable, massive and weakly 
interacting~\cite{Goldberg:1983nd,RelicCP}.  
SUSY models can also provide new sources of CP violation~\cite{Haber:1997if}.
In the Minimal Supersymmetric Standard Model (MSSM)~\cite{mssm},  
the complex parameters are conventionally chosen to be the Higgsino mass 
parameter $\mu$, the  ${\rm U(1)}$ and ${\rm SU(3)}$  gaugino mass parameters $M_1$ and 
$M_3$, respectively, and the trilinear scalar coupling parameters $A_f$ of the third 
generation sfermions ($f=b,t,\tau$),
\begin{eqnarray}
\label{eq:phases}
	\mu = |\mu| e^{i \phi_\mu}, \quad  
	M_1 = |M_1| e^{i \phi_{1}}, \quad
	M_3 = |M_3| e^{i \phi_3},   \quad
	A_f = |A_f| e^{i \phi_{A_f}}.
\end{eqnarray}
These phases contribute to the electric dipole moments (EDMs),
in particular to those of  Thallium~\cite{Regan:2002ta}, Mercury~\cite{Griffith:2009zz},
the neutron~\cite{Baker:2006ts}, and the deuteron~\cite{Semertzidis:2003iq},
which can be beyond their current experimental upper 
bounds~\cite{Regan:2002ta, Baker:2006ts,Griffith:2009zz,Semertzidis:2003iq,Amsler:2008zz}.
The experimental limits generally restrict the CP phases to be smaller than $\pi/10$, 
in particular the phase $\phi_\mu$~\cite{Choi:2004rf,Barger:2001nu}. 
However, the extent to which the EDMs can constrain the SUSY  phases strongly depends on 
the considered model and its 
parameters~\cite{CP-Problem,EDM,Barger:2001nu,Bartl:2003ju,Ellis:2008zy,Choi:2004rf,Deppisch:2009nj}.

\medskip

As shown for example in Ref~\cite{Deppisch:2009nj},  due to cancellations among different 
contributions to the EDMs, only isolated points in the CP phase space can give large 
CP-violating signals  at the LHC. 
It is important to search for these signals, since  the cancellations could be a 
consequence of a deeper model that correlates the phases. In addition, 
the existing EDM bounds can also be fulfilled by including lepton flavor violating 
couplings in the slepton sector~\cite{Bartl:2003ju}. This is important when considering 
for example SUSY Seesaw models, where CP violation in the slepton sector is connected 
to the neutrino sector and Leptogenesis~\cite{Leptogen,reviewLeptogen}. 
\medskip

Thus measurements of SUSY CP observables outside the low energy EDM sector are necessary 
to independently determine or constrain the phases.   
In particular,  the phases of the trilinear scalar coupling parameters $A_f$ have a 
significant impact on the MSSM Higgs sector~\cite{Pilaftsis}. Loop effects, dominantly 
mediated by third generation squarks, can generate large CP-violating scalar-pseudoscalar 
transitions among the neutral Higgs bosons~\cite{HA,Choi:2004kq}. As a result, the lightest Higgs 
boson with a mass of order $10$~GeV or $45$~GeV \cite{Carena:2000ks} cannot be excluded 
by measurements at LEP~\cite{LEPbounds}. The fundamental properties and the 
phenomenology of CP-violating neutral Higgs boson mixings have been investigated in 
detail in the literature~\cite{Hreview,Accomando:2006ga}.

\medskip

The phases can also drastically change other SUSY particle masses, their  cross sections, 
branching ratios~\cite{Bartl:2003he,Bartl:2002uy,Alan:2007rp,Rolbiecki:2009hk}, and 
longitudinal polarizations of final fermions~\cite{Gajdosik:2004ed}. Although such 
CP-even observables can be very sensitive to the CP phases (the observables can change by an 
order of magnitude and more), CP-odd (T-odd) observables have to be measured for a 
direct evidence of CP violation. CP-odd observables are, for example, rate asymmetries 
of cross sections, distributions, and partial decay widths~\cite{otherCPodd}. However, 
these observables require the presence of absorptive phases, e.g. from loops. Thus 
they usually do not exceed the size of $10\%$, unless they are resonantly
enhanced~\cite{Choi:2004kq,Accomando:2006ga, Pilaftsis:1997dr,Nagashima:2009gm}.

\medskip

Larger CP asymmetries in particle decay chains can be obtained with triple products of 
final particle  momenta~\cite{tripleprods}. They already appear at tree level due to 
spin correlations. Triple product asymmetries have been intensively studied in the 
production and decay of neutralinos~\cite{NEUT2,NEUT3,Bartl:2003ck,
AguilarSaavedra:2004hu,Kittel:2004rp} and charginos~\cite{Kittel:2004rp,Kittel:2004kd,CHAR2,CHAR3} 
at the ILC~\cite{ILC}, also using transversely polarized beams~\cite{Trans}. 
At the LHC~\cite{LHC}, triple product asymmetries have been studied for the decays of 
neutralinos~\cite{Bartl:2003ck,Langacker:2007ur,MoortgatPick:2009jy}, 
stops~\cite{Bartl:2004jr,Ellis:2008hq,Deppisch:2009nj}, 
and sbottoms~\cite{Bartl:2006hh}.
For recent reviews, see Ref.~\cite{CPreview}.

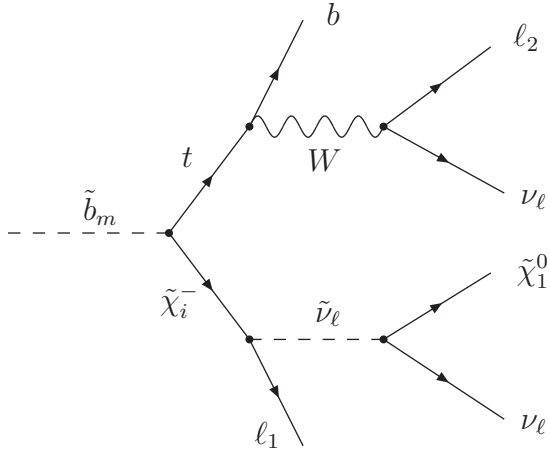
\begin{figure}[t]
\scalebox{1}{
\begin{picture}(10,5)(-3.4,0)
	\DashLine(-10,80)(50,80){5}
	\Vertex(50,80){1.5}
	\ArrowLine(50,80)(80,120)
	\Vertex(80,120){1.5}
	\ArrowLine(80,120)(100,160)
	\Photon(80,120)(130,120){4}{4}
	\Vertex(130,120){1.5}
	\ArrowLine(130,120)(170,150)
	\ArrowLine(130,120)(175,95)
	\ArrowLine(50,80)(80,40)
	\Vertex(80,40){1.5}
	\ArrowLine(80,40)(100,0)
	\DashLine(80,40)(130,40){5}
	\Vertex(130,40){1.5}
	\ArrowLine(130,40)(170,65)
	\ArrowLine(130,40)(175,10)
	\put( 0.5,3.0){ $\tilde b_m $}
	\put( 1.51,1.8){ $\tilde\chi_i^- $}
	\put( 2.75,0.03){ $\ell_1$}
	\put(3.55,1.65){ $\tilde\nu_\ell$}
	\put(6.2,2.2){ $\tilde\chi_1^0$}
	\put(6.25,0.22){ $ \nu_\ell$}
	\put( 1.8,3.7){ $t$}
	\put(3.45,3.64){ $W$}
	\put( 3.7,5.58){ $b$}
	\put(6.25, 3.2){ $\nu_\ell $}
	\put(6.16,5.3){ $\ell_2$}
\end{picture}}
\caption{Schematic picture of bottom squark decay.}
\label{Fig:decaySbottom}
\end{figure}

\medskip

In this paper, we thus study CP asymmetries in two-body decays of a sbottom,
\begin{eqnarray}
\label{eq:decaySbottom}
	\tilde b_m \to t + \tilde\chi^-_i; \quad m=1,2; \quad i=1,2;
\end{eqnarray}
followed by the subsequent two-body decay  of the chargino,
\begin{eqnarray}
\label{eq:decayChi}
	\tilde\chi^-_i \to \ell_1^- + \tilde\nu_\ell^\ast, \quad  \ell= e,\mu,
\end{eqnarray}
with the invisible sneutrino decay
$\tilde\nu_\ell^\ast \to \tilde\chi^0_1 \, \bar\nu_\ell$,
and the subsequent top decay
\begin{eqnarray}
\label{eq:decayTop}
	t \to b + W^+; \quad 
	W^+ \to \nu_\ell + \ell_2^+; \qquad \ell = e,\mu,
\end{eqnarray}
see Figure~\ref{Fig:decaySbottom} for a schematic picture of the entire sbottom 
decay chain. 
The CP-sensitive  spin-spin correlations of the sbottom decay allow us to probe 
the phase of the coupling parameter $A_{b}$, and the phase of the
higgsiono mass parameter $\mu$.
The phases of the trilinear scalar coupling parameters of the third 
generation sfermions are rather unconstrained by the 
EDMs~\cite{Ellis:2008zy,Choi:2004rf}. Therefore it is appealing to study 
CP asymmetries in squark decays at high energy 
colliders like the LHC~\cite{LHC} or ILC~\cite{ILC}.
The third generation sfermions also have a rich phenomenology due to a 
sizeable mixing of left and right states. 

\medskip

CP asymmetries based on triple products have already been studied in two-body decays of 
sbottoms~\cite{Bartl:2006hh}, however in their rest frame only.  
At colliders like the LHC, the particles are highly boosted, which will generally 
reduce the triple product 
asymmetries~\cite{Ellis:2008hq,Langacker:2007ur,Deppisch:2009nj,MoortgatPick:2009jy}. 
We thus will also include the sbottom production at the LHC. 
In addition, we will also study asymmetries which base on epsilon products.
Those asymmetries are boost invariant, and thus provide the largest possible asymmetries. 
We calculate the amplitude squared for the entire sbottom decay in the 
spin-density matrix formalism~\cite{Haber:1994pe}.  The compact form of the amplitude 
squared allows us to identify the optimal CP observables for sbottom decays, an
asymmetry, that has not been studied in Ref.~\cite{Bartl:2006hh}.

\medskip

In Section~\ref{sec:TripleProducts}, we identify the 
CP-sensitive parts in the amplitude squared, define the CP asymmetries in bottom squark 
decays, and discuss their dependence on the complex sbottom-top-chargino couplings. 
We analyze their MSSM parameter dependence in a SUSY benchmark scenario.
In Section~\ref{sec:AsymmetryLHC}, we discuss sbottom production and their boost 
distribution at the LHC. We give lower bounds on the required LHC luminosities to 
observe the asymmetries over their statistical fluctuations.  
We summarize and give our conclusions in Section~\ref{sec:Conclusion}. 
In the Appendix, we review sbottom mixing  with complex parameters,
 give the phase space, and calculate the sbottom decay 
amplitudes in the spin-density matrix formalism.

\section{CP asymmetries in bottom squark decays}
\label{sec:TripleProducts}

In this Section, we identify the CP-sensitive parts in the amplitude squared of the 
entire two-body decay chain of the bottom squark, see Eqs.~(\ref{eq:decaySbottom})-(\ref{eq:decayTop})
 and Figure~\ref{Fig:decaySbottom}. In order to probe these parts, we define
CP asymmetries of epsilon and triple  products of the particle momenta. 
Explicit expressions for the squared amplitude, Lagrangians, couplings, and phase-space 
elements are summarized in the Appendix. 

\subsection{T-odd products}
\label{sec:CPterms}

The amplitude squared $|T|^2$ for the sbottom decay chain, see Figure~\ref{Fig:decaySbottom},
can be decomposed into contributions from the top spin correlations, the chargino 
spin correlations, the top-chargino spin-spin correlations, and 
an unpolarized part, see Eq.~(\ref{eq:matrixelement2}).
Since we have a two-body decay of a scalar particle,
a CP-sensitive part can only originate from  the spin-spin correlations.
It is given by the last summand in Eq.~(\ref{CPterm}), which is 
proportional to 
\begin{equation}
	|T|^2 \supset
	{\rm Im}\{ l^{\tilde b}_{mi} (k^{\tilde b}_{mi})^\ast \}
	[p_{\tilde b},p_t,p_{\ell_1},p_{\ell_2}], \quad m,i=1,2.
\label{CPterm2}
\end{equation}
The left and right couplings  $k^{\tilde b}_{mi}$ and $l^{\tilde b}_{mi}$, respectively, 
are defined through the $\tilde b_m$--$t$--$\tilde\chi^\pm_i$ 
Lagrangian~\cite{Bartl:2004jr}, 
\begin{eqnarray}
{\scr L}_{t \tilde b\tilde\chi^+}
= g\,\bar t\,(l_{mi}^{\tilde b}\,P_R + k_{mi}^{\tilde b}\,P_L)
            \,\tilde\chi_i^+\,\tilde b_m + {\rm h.c.} ,
\end{eqnarray}
see Appendix~\ref{Lagrangians and couplings}.
These couplings depend on the mixing in the sbottom and chargino sector, and thus 
on the CP phases $\phi_{A_b}$ and $\phi_{\mu}$. The imaginary part of the coupling 
product, ${\rm Im}\{ l^{\tilde b}_{mi} (k^{\tilde b}_{mi})^\ast \}$, 
in Eq.~(\ref{CPterm2}) is multiplied by a T-odd epsilon product ${\mathcal E}$, 
for which we  use the short hand notation 
\begin{equation}
	{\mathcal E}\equiv
	[p_{\tilde b},p_t,p_{\ell_1},p_{\ell_2}]\equiv
	\varepsilon_{\mu\nu\alpha\beta}~
	p_{\tilde b}^{\mu}~p_t^{\nu}~p_{\ell_1}^{\alpha}~p_{\ell_2}^{\beta},
\label{eq:epsilon}                
\end{equation}
with the convention $\varepsilon_{0123}=1$. Since each of the spatial components of the 
four-momenta changes sign under a naive time transformation, $t \to -t$, this product 
is T-odd. Due to CPT invariance, T-odd products are related to CP-odd observables.



\subsection{T-odd asymmetries}
\label{sec:TOddProducts}
The task is to define an observable, that projects out the CP-sensitive
part of the spin-spin correlation term from the amplitude squared. 
This can be achieved by defining for the T-odd product ${\mathcal E}$, 
Eq.~(\ref{eq:epsilon}), the T-odd asymmetry of the partial 
sbottom decay width $\Gamma$~\cite{Deppisch:2009nj},
\begin{equation}
	{\mathcal A}=
	\frac{\Gamma({\mathcal E}>0)-\Gamma({\mathcal E}<0)}
	     {\Gamma({\mathcal E}>0)+\Gamma({\mathcal E}<0)}
	=
	\frac{\int{\rm Sign}[{\mathcal E}] |T|^2 d{\rm Lips}}
	     {\int |T|^2 d{\rm Lips}},
\label{eq:Toddasym}
\end{equation}
with the amplitude squared $|T|^2$, 
and the Lorentz invariant phase-space element $d{\rm Lips}$, 
such that $\int |T|^2 d{\rm Lips}/( 2m_{\tilde b})=\Gamma$. 
The T-odd asymmetry is also CP-odd, if absorptive phases (from higher order final-state
interactions or finite-width effects) can be neglected~\cite{tripleprods}.

\medskip

In general, largest asymmetries are obtained by using the epsilon product 
${\mathcal E}$, see Eqs.~(\ref{CPterm2}) and (\ref{eq:epsilon}), that matches the 
kinematic dependence of the CP-sensitive terms in the amplitude squared. In the literature, 
this technique is sometimes referred to \emph{optimal observables}~\cite{optimal}.
%
Other possible combinations\footnote{
       Note  that momenta from both the decay products of the chargino 
       \emph{and} the top have to be included 
       to obtain non-vanishing asymmetries~\cite{Bartl:2006hh}. 
       Otherwise the CP-sensitive top-chargino spin-spin correlations are lost. 
       If for example momenta from the top decay are not taken into account, only 
       CP asymmetries from the chargino decay can be obtained.
       In addition,  a three-body decay is required,
        (or a two-body decay via an on-shell $W$ boson~\cite{Kittel:2004kd})
       to probe the  phases $\phi_{\mu}$,  $\phi_{1}$ of the 
       chargino/neutralino system. 
       The asymmetries are then of the order of $10\%$, which is typical for 
       chargino three-body decays~\cite{CHAR3}, and also neutralino three-body 
       decays~\cite{NEUT3,Ellis:2008hq, AguilarSaavedra:2004hu, 
       Langacker:2007ur,MoortgatPick:2009jy}.
                  }
of momenta, ${\mathcal E}=[p_{\tilde b},p_b,p_{\ell_1},p_{\ell_2}]$,
or ${\mathcal E}=[p_{\tilde b},p_t,p_{b},p_{\ell_1}]$,
lead to smaller asymmetries, see Ref.~\cite{Bartl:2006hh},
and Fig.~\ref{fig:BoostedAsymmetry}.

\medskip

Triple products of three spatial momenta can also be used to define 
asymmetries~\cite{tripleprods,Bartl:2006hh}. 
In the sbottom rest frame, $p_{\tilde b}^\mu = (m_{\tilde b}, \mathbf{0})$, 
the epsilon product is
\begin{equation}
	[p_{\tilde b},p_t,p_{\ell_1},p_{\ell_2}] =
	m_{\tilde b} \; \,\mathbf{p}_t\cdot 
        ( \mathbf{p}_{\ell_1 } \times 	\mathbf{p}_{\ell_2}) \equiv 
	m_{\tilde b} \, {\mathcal T}.
\label{tripleterm}
\end{equation}
That triple product will give the largest asymmetries in the sbottom rest frame, and 
the other combinations of momenta for ${\mathcal T}$
lead to smaller asymmetries. 

\medskip
Note that the asymmetries of an epsilon product 
${\mathcal E}$ are by construction Lorentz invariant whereas those constructed with
 a triple product ${\mathcal T}$ are not~\cite{Langacker:2007ur,Deppisch:2009nj}. 
The triple product asymmetries will therefore depend on the sbottom boost, 
$\beta_{\tilde b} = |{\mathbf p}_{\tilde b}|/E_{\tilde b}$, and are generally reduced 
if not evaluated in the sbottom rest frame. We will discuss the impact 
of the sbottom boost on the asymmetries at the LHC in Section~\ref{sec:AsymmetryLHC}.

\newpage
\subsection{Parameter and phase dependence}
\label{sec:PhaseDependence}

%
\begin{table}[t]
\renewcommand{\arraystretch}{1.6}
\caption{
         MSSM scenario. The mass dimension parameters are given in GeV.
\label{tab:ReferenceScenario}}
\begin{center}
\begin{tabular}{cccccccc} 
\hline
  \multicolumn{1}{c}{$ |\mu| $} 
& \multicolumn{1}{c}{$ M_{2} $}
& \multicolumn{1}{c}{$ M_{\tilde D} $}
& \multicolumn{1}{c}{$ M_{\tilde Q} $}
& \multicolumn{1}{c}{$ \tan\beta $}
& \multicolumn{1}{c}{$ A_b $}  
& \multicolumn{1}{c}{$ \phi_{A_b} $}
& \multicolumn{1}{c}{$ \phi_{\mu}=\phi_{M_1} $}
\\\hline
  \multicolumn{1}{c}{$ 200 $} 
& \multicolumn{1}{c}{$ 250 $}
& \multicolumn{1}{c}{$ 400 $}
& \multicolumn{1}{c}{$ 420 $}  
& \multicolumn{1}{c}{$ 5 $} 
& \multicolumn{1}{c}{$ 1200 $}
& \multicolumn{1}{c}{$ \frac{1}{5}\,\pi $}
& \multicolumn{1}{c}{$ 0 $}
\\\hline
\end{tabular}
\end{center}
\renewcommand{\arraystretch}{1.0}
\end{table}

In order to analyze the phase and MSSM parameter
dependence of the asymmetry ${\mathcal A}$, 
Eq.~(\ref{eq:Toddasym}), we insert the explicit form of the amplitude squared 
in the spin-density formalism. As shown in Appendix~\ref{Sbottom decay width}, we obtain
\begin{equation}
	{\mathcal A}=  \eta \;
	\frac{\int {\rm Sign}({\mathcal E})
	(p_b\cdot p_{\nu_\ell})\,
	[p_{\tilde b},p_t,p_{\ell_1},p_{\ell_2}]~
	d{\rm Lips}  }{(p_{\chi_i^\pm}\cdot p_{\ell_1})\int
	(p_t\cdot p_{\ell_2})
	(p_b\cdot p_{\nu_\ell})~
	d{\rm Lips} },
\label{eq:Adependence2}
\end{equation}
with $ (p_{\chi_i^\pm}\cdot p_{\ell_1})= (m_{\chi_i^\pm}^2- m_{\tilde\nu_\ell }^2)/2$,
and the coupling function
\begin{equation}
\eta = \frac{{\rm Im}\{ l^{\tilde b}_{mi} (k^{\tilde b}_{mi})^\ast \}}
      { \frac{1}{2}\left(|l^{\tilde b}_{mi}|^2 + |k^{\tilde b}_{mi}|^2\right)
     \frac{m_{\tilde b}^2 - m_{\chi_i^\pm}^2 - m_t^2}
    {2 m_{\chi_i^\pm} m_t}
     - {\rm Re}\{ l^{\tilde b}_{mi} (k^{\tilde b}_{mi})^\ast \}}.
\label{eq:couplingfunct}
\end{equation}
The asymmetry can thus be separated into a kinematical part and the effective coupling 
factor $\eta$. That factor is approximately independent of the particle masses, but governs 
the main dependence on the CP phases, and on the parameters of the 
sbottom-top-chargino couplings $l^{\tilde b}_{mi}$, $k^{\tilde b}_{mi}$.
%
%
To qualitatively understand this dependence, 
we expand~\cite{Bartl:2006hh}
\begin{equation}
{\rm Im}\{ l^{\tilde b}_{mi} (k^{\tilde b}_{mi})^\ast \}
= c_m\,Y_t \,{\rm Im}\{U_{i1}^\ast V_{i2}^\ast  \}
-\frac{1}{2}\,d_m Y_t\, Y_b\sin( 2\theta_{\tilde b})
{\rm Im}\{U_{i2}^\ast V_{i2}^\ast  e^{-i\phi_{\tilde b}}\},
\label{expansion}
\end{equation}
with the  Yukawa couplings $Y_t$ and $Y_b$,  see Eq.~(\ref{eq:ytb}),
the short hand notation 
$c_1=\cos^2\theta_{\tilde b}$, $c_2=\sin^2\theta_{\tilde b}$, $d_1=1$, $d_2=-1$, 
the sbottom mixing angle $\theta_{\tilde b}$, Eq.~(\ref{eq:thsbottom}), 
and the CP phase  
$\phi_{\tilde b} =  \arg\lbrack A_b-\mu^\ast\tan\beta\rbrack$,
Eq.~(\ref{eq:phib}), of the sbottom system. 
The imaginary part of the products of the sbottom-top-chargino 
couplings ${\rm Im}\{ l^{\tilde b}_{mi} (k^{\tilde b}_{mi})^\ast \}$
 is sensitive to the  phases  $\phi_\mu$ and $\phi_{A_b}$,
and can be large due to the Yukawa couplings $Y_t, Y_b$.
For  $\phi_\mu=0$ or $\pi$, the chargino diagonalization matrices $U,V$ are real,
and~\cite{Bartl:2006hh}
\begin{equation}
{\rm Im}\{ l^{\tilde b}_{mi} (k^{\tilde b}_{mi})^\ast \}
\;\propto\;
\sin( 2\theta_{\tilde b})\sin(\phi_{\tilde b}),
\end{equation}
%
In particular, for $\phi_\mu=0$ or $\pi$,
  ${\rm Im}\{ l^{\tilde b}_{mi} (k^{\tilde b}_{mi})^\ast \} $ 
shows a $\sin(\phi_{ A_b})$ behaviour,
and is maximal at $\sin(\phi_{A_b})\approx \pi/2,3\pi/2$.
However, the maxima of the asymmetries correspond to the maxima of $\eta,$
which are shiftet away from $\phi_{A_b} = \pi/2,3\pi/2$. This is due to
the influence of the denominator of $\eta$, see Eq.~(\ref{eq:couplingfunct2}), 
which has a cosine-like dependence on $\phi_{A_b}$, and 
$\eta$ will be approximately maximal for 
$ |l^{\tilde b}_{mi}| \approx |k^{\tilde b}_{mi}|$.

\begin{table}[t]
\renewcommand{\arraystretch}{1.6}
\caption{ 
       SUSY particle masses for the benchmark scenario of 
       Table~\ref{tab:ReferenceScenario} ($\ell = e,\mu$).
         \label{tab:ReferenceScenarioDerived}}
\begin{center}
      \begin{tabular}{|c|c|c|c|}
\hline
$   m_{\tilde b_1}  =   400~{\rm GeV}$ &
$   m_{\tilde b_2}  =   424~{\rm GeV}$ &
$   m_{\chi^\pm_1}  =   158~{\rm GeV}$ &
$   m_{\chi^\pm_2}  =   301~{\rm GeV}$ 
\\ \hline
$   m_{\chi^0_1}  =   108~{\rm GeV}$ &
$   m_{\chi^0_2}  =   172~{\rm GeV}$ &
$   m_{\chi^0_3}  =   206~{\rm GeV}$ &
$   m_{\chi^0_4}  =   302~{\rm GeV}$ 
\\ \hline
$    m_{\tilde\nu_\ell}  =   137~{\rm GeV}$ &
$    m_{\tilde\tau_1}    =   152~{\rm GeV}$ &
$    m_{\tilde\ell_R}    =   156~{\rm GeV}$ & 
$    m_{\tilde\ell_L}    =   157~{\rm GeV}$  
\\ \hline
\end{tabular}
\end{center}
\renewcommand{\arraystretch}{1.0}
\end{table}

\subsection{Numerical analysis}
\label{sec:Numericalanalysis}
%
To quantitatively study the asymmetry ${\mathcal A}$, Eq.~(\ref{eq:Toddasym}), 
we analyze the behaviour of the coupling factor $\eta$ in more detail,
and define a benchmark scenario in Table~\ref{tab:ReferenceScenario}. 
We give the relevant resulting SUSY  masses in Table~\ref{tab:ReferenceScenarioDerived}.
 We fix the soft-breaking parameters in the slepton sector by
$M_{\tilde E}^{\tilde \ell}=M_{\tilde L}^{\tilde \ell} =150$~GeV
for $\ell=e, \mu,\tau$, to enable the subsequent chargino decay 
$\tilde\chi^\pm_1 \to \ell_1^\pm \tilde\nu_\ell^{(\ast)}$.
We take stau mixing into account, and fix
the trilinear scalar coupling parameter $A_\tau=250$~GeV.  
Since its phase does not contribute to the CP asymmetry, we set
it to $\phi_{A_\tau}=0$, as well as $\phi_1=0$, which is the
phase of the U(1) gaugino mass parameter $M_1$.
In order to reduce the number of parameters, we further use the GUT 
inspired\footnote{Note that this choice also can significantly constrain 
                  the neutralino sector~\cite{Neutpaper}.
} 
relation $|M_1|=5/3\,M_2\tan^2 \theta_w$. 

\medskip

We will study the decay of the lighter sbottom into the lightest chargino,
$\tilde b_1 \to t \tilde\chi^\pm_1$,
which gives the dominant contribution to the asymmetries.
For our reference scenario 
as defined in Table~\ref{tab:ReferenceScenario},
the corresponding branching ratios are
${\rm BR}(\tilde b_1 \to t \tilde\chi^\pm_1)=19\%$ and
${\rm BR}(\tilde\chi^\pm_1 \to e^\pm \tilde\nu_e)=33\%$.
Other decay channels yield much smaller asymmetries for the scenario 
as defined in Table~\ref{tab:ReferenceScenario}.
Further, we will focus on the largest (Lorentz invariant) asymmetry 
for the optimal T-odd product 
${\mathcal E} =	[p_{\tilde b},p_t,p_{\ell_1},p_{\ell_2}]$,
see Eq.~(\ref{eq:epsilon}).
Note that in the sbottom rest frame, this asymmetry is equivalent to
the triple product asymmetry with
${\mathcal T} = {\mathbf p}_t \cdot (\mathbf{p}_{\ell_1}\times\mathbf{p}_{\ell_2})$,
see Eq.~(\ref{tripleterm}).
We use the short hand notation  $\mathcal{A}(t\ell_1\ell_2)$
for the asymmetry, to indicate the momenta used for the triple product.


\begin{figure}[t!]
\centering
\includegraphics[clip,width=0.495\textwidth]{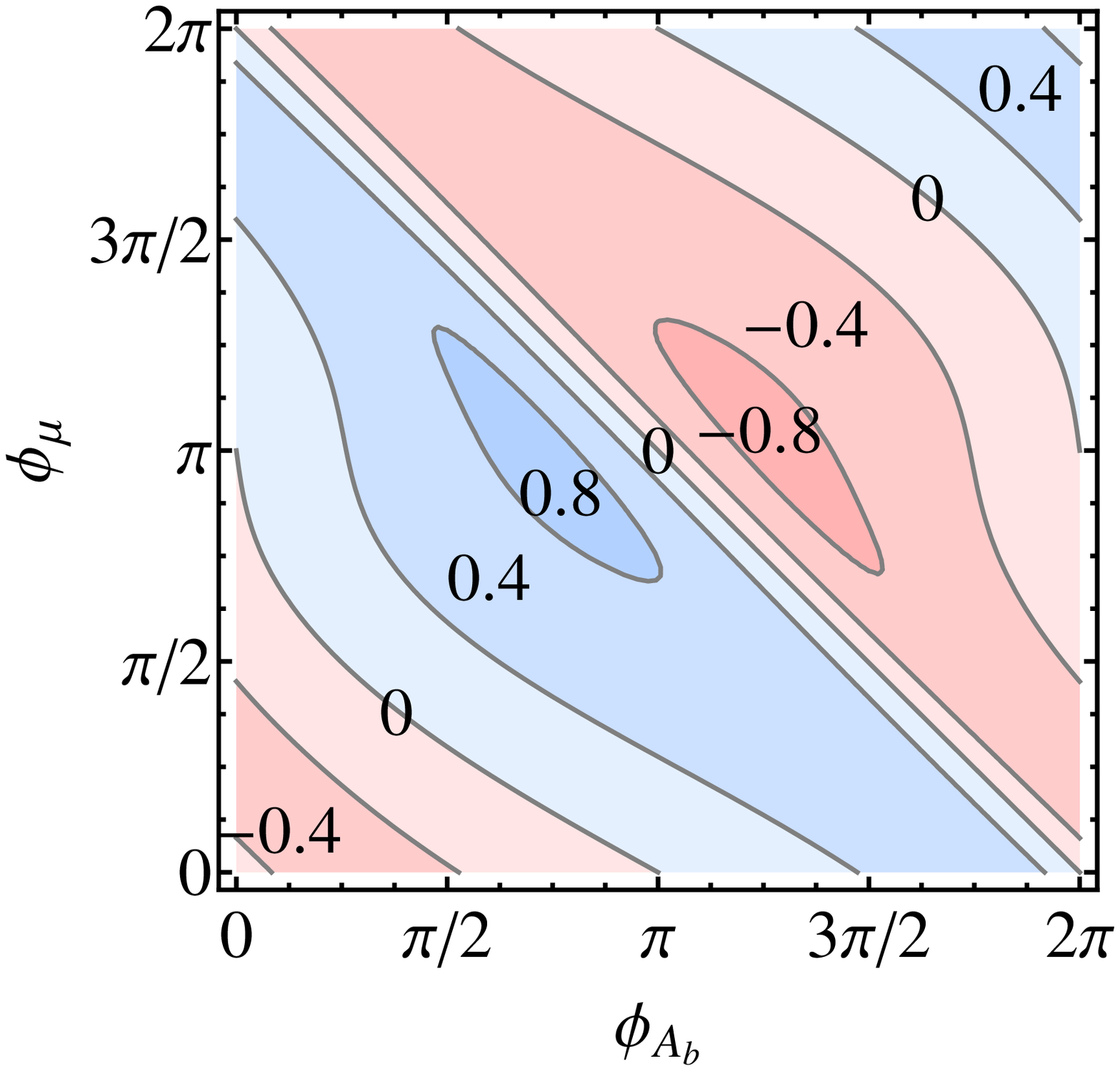}
\includegraphics[clip,width=0.495\textwidth]{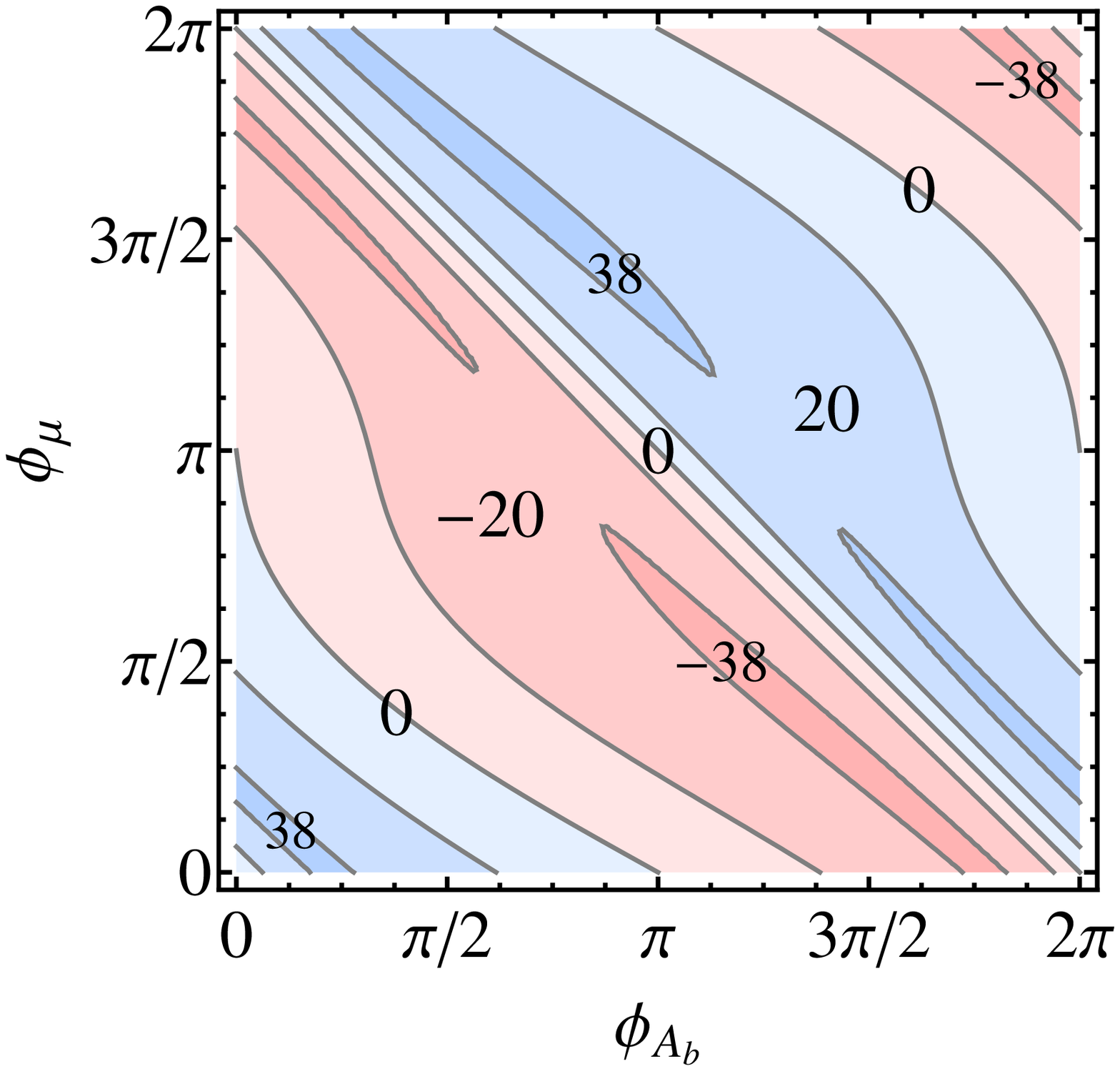}
\caption{
         Phase dependence of the  CP-odd coupling   
         factor $\eta$, Eq.~(\ref{eq:couplingfunct}),
         for sbottom decay  $\tilde b_1 \to t \tilde\chi^-_1$ (left),
         and the corresponding CP asymmetry  ${\mathcal A}(t\ell_1\ell_2)$, 
         Eq.~(\ref{eq:Toddasym}), in percent (right), for the subsequent two-body decay
         chain  $\tilde\chi^-_1 \to \ell_1^- \,\tilde\nu_\ell^{\ast}$,
         and $t\to b\,\nu_\ell\,\ell_2^+ $,
         see Figure~\ref{Fig:decaySbottom}, in the $\tilde b_1$ rest frame.
         The SUSY parameters are given in Table~\ref{tab:ReferenceScenario}.
}
\label{fig:asym_eta_phases}
\end{figure}
\medskip

In Figure~\ref{fig:asym_eta_phases}, we show 
the dependence of the coupling factor $\eta$,  Eq.~(\ref{eq:couplingfunct}),
and its corresponding asymmetry ${\mathcal A}(t\ell_1\ell_2)$, Eq.~(\ref{eq:Adependence2}),
on the two CP phases $\phi_{A_b}$ and $\phi_{\mu}$, for the SUSY scenario of 
Table~\ref{tab:ReferenceScenario}. 
We clearly see that the maxima of $\mathcal{A}(t\ell_1\ell_2) \approx\pm 40\%$ 
are not necessarily obtained for maximal CP phases,
$\phi_{A_b, \mu} = \pi/2,3\pi/2$.
The reason is that the phase dependence of 
$\mathcal{A}$ is almost governed by the coupling factor $\eta$, 
shown in the left panel of Figure~\ref{fig:asym_eta_phases}.
Thus the asymmetry can be sizeable even for small values of the phases, favored by 
EDM constraints, which in particular  constrain $\phi_{\mu}$.
For example, the asymmetry has a maximum of 
$\mathcal{A}(t\ell_1\ell_2)\approx\pm 40\%$
at small $\phi_\mu\approx \pm 0.2\pi$, for $\phi_{A_b}=0$.
The positions of the maxima will remain when including the effects of the 
sbottom boost, as we will discuss in Section~\ref{sec:AsymmetryLab}. 
Figure~\ref{fig:asym_eta_phases} motivates the choice 
$\phi_{A_b}=  0.2\pi$,  $\phi_\mu=0$, in our 
benchmark scenario, Table~\ref{tab:ReferenceScenario}.


\begin{figure}[t!]
\centering
\includegraphics[clip,width=0.495\textwidth]{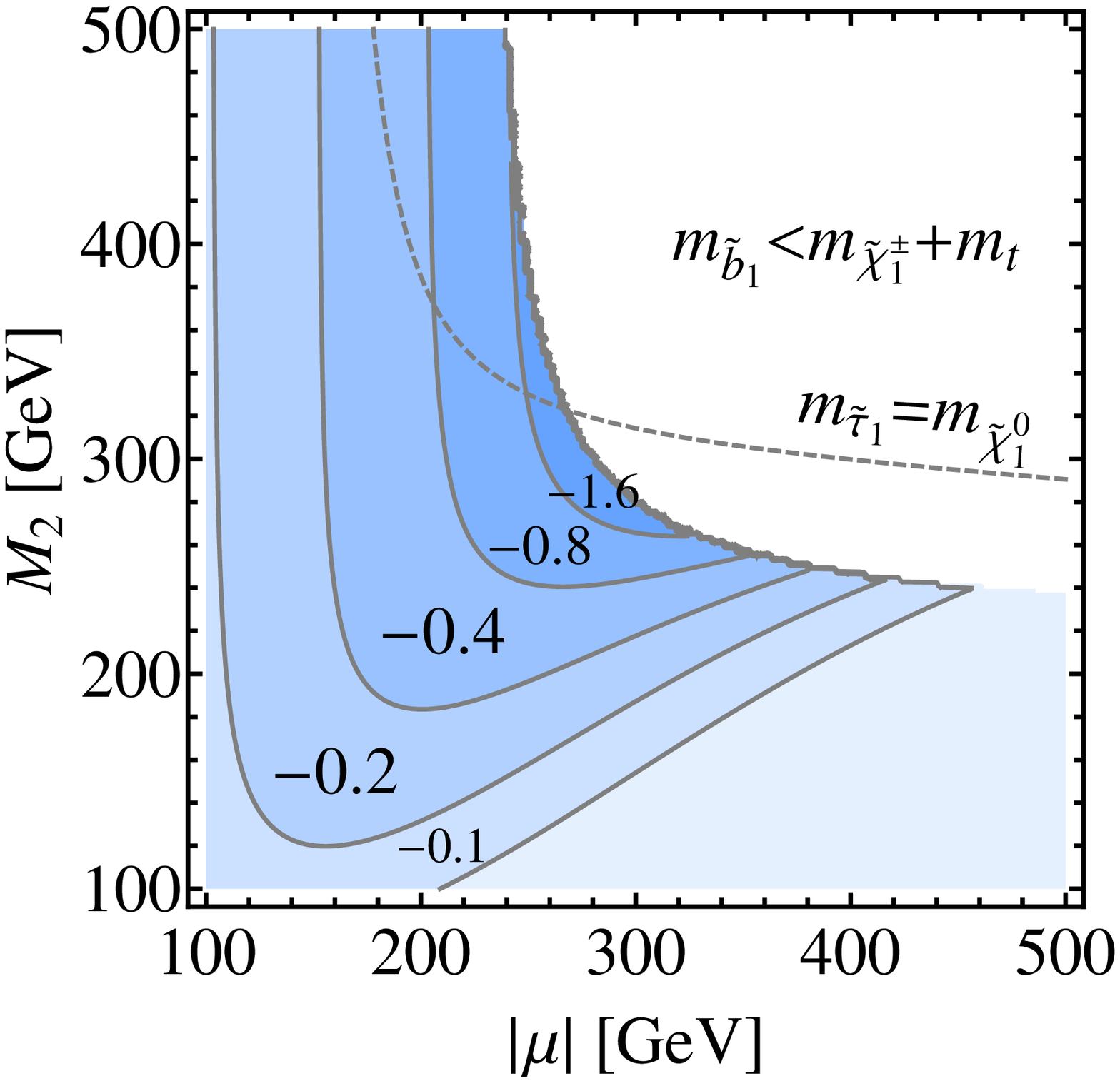}
\includegraphics[clip,width=0.495\textwidth]{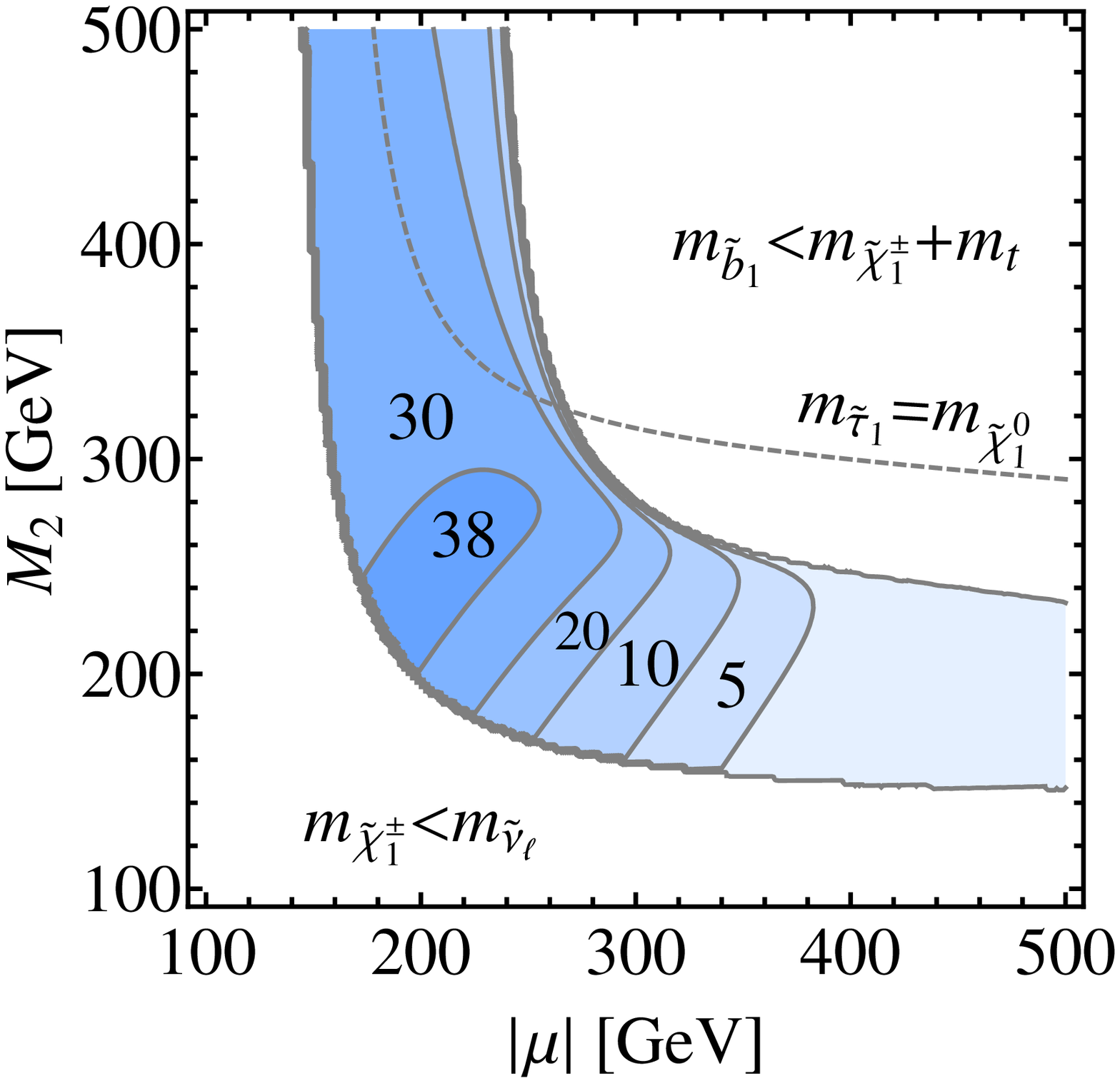}
\caption{
         Contour lines in the $\mu$--$M_2$ plane of the CP-odd coupling   
         factor $\eta$, Eq.~(\ref{eq:couplingfunct}),
         for sbottom decay  $\tilde b_1 \to t \tilde\chi^-_1$ (left),
         and the corresponding CP asymmetry  ${\mathcal A}(t\ell_1\ell_2)$, 
         Eq.~(\ref{eq:Toddasym}), in percent (right), for the subsequent two-body decay
         chain  $\tilde\chi^-_1 \to \ell_1^- \,\tilde\nu_\ell^{\ast}$,
         and $t\to b\,\nu_\ell\,\ell_2^+ $,
         see Figure~\ref{Fig:decaySbottom}, in the sbottom rest frame.
         The SUSY parameters are given in Table~\ref{tab:ReferenceScenario}.
         The area  above the contour lines is kinematically forbidden by
         $m_{\tilde b_1} < m_{\chi_1^\pm} +m_t$,
         the area  below the contour lines of the asymmetry (right)
         is  forbidden by $m_{\chi_1^\pm}< m_{\tilde\nu_\ell} $.
         The area above the dashed line is excluded by 
         $m_{\tilde\tau_1} <m_{\chi_1^0} $.
}
\label{fig:asym_eta_mum2}
\end{figure}

\medskip

\newpage
In Figure~\ref{fig:asym_eta_mum2}, we present the  dependence of the 
asymmetry (left) and the  coupling factor $\eta$ (right)
on the chargino mixing, which is mainly determined by
the higgsino and gaugino parameters $\mu$ and $M_2$, respectively.
We find large values of the coupling factor $\eta$ 
and the asymmetry for mixed and higgsino-like charginos.
The kinematical boundaries for the sbottom decay  
$\tilde b_1 \to t \tilde\chi^-_1$, and the chargino
decay $\tilde\chi^-_1 \to \ell_1^- \,\tilde\nu_\ell^{\ast}$
are indicated in Figure~\ref{fig:asym_eta_mum2}.


\begin{figure}[t!]
\centering
\includegraphics[clip,width=0.495\textwidth]{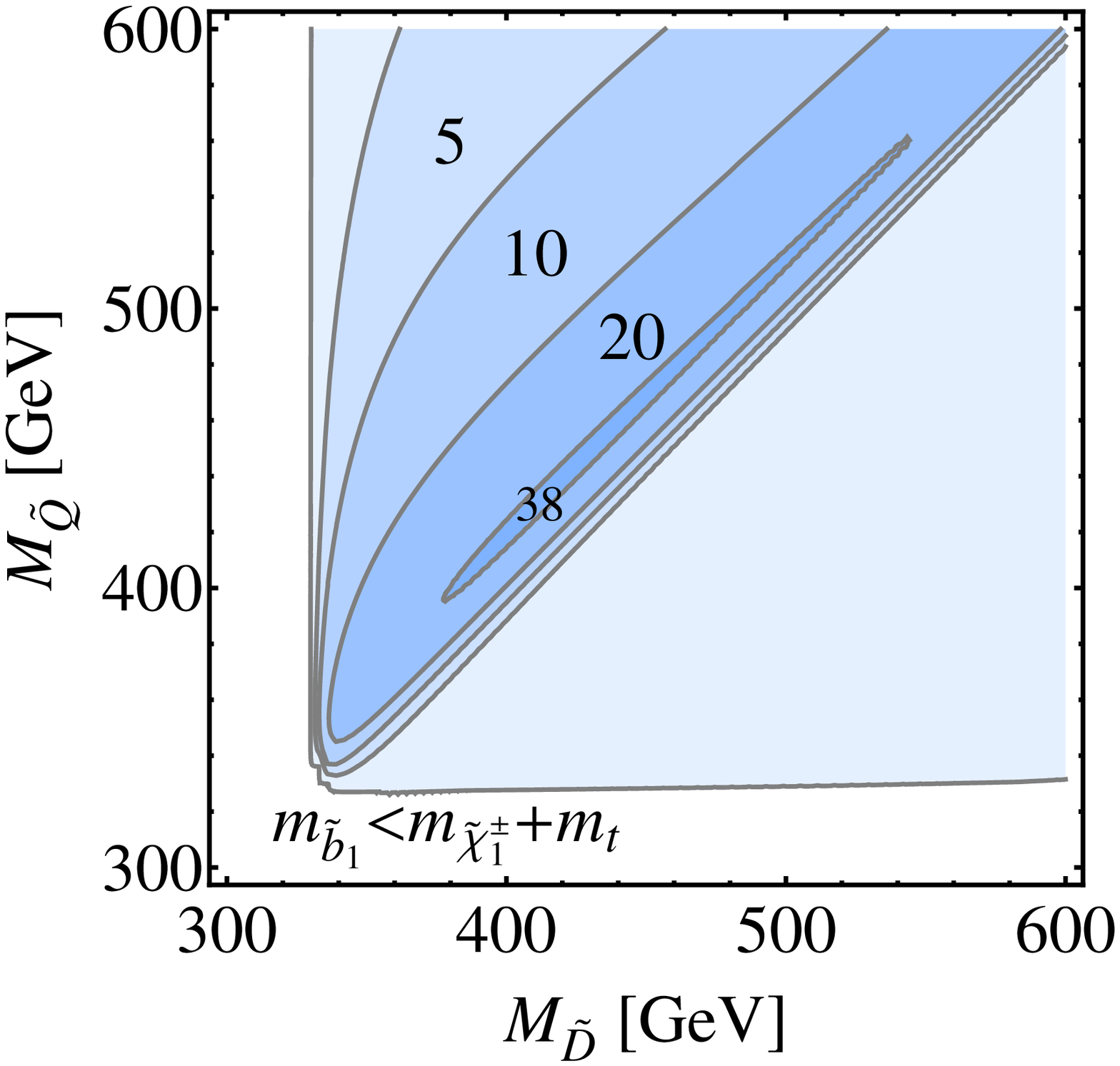}
\includegraphics[clip,width=0.473\textwidth]{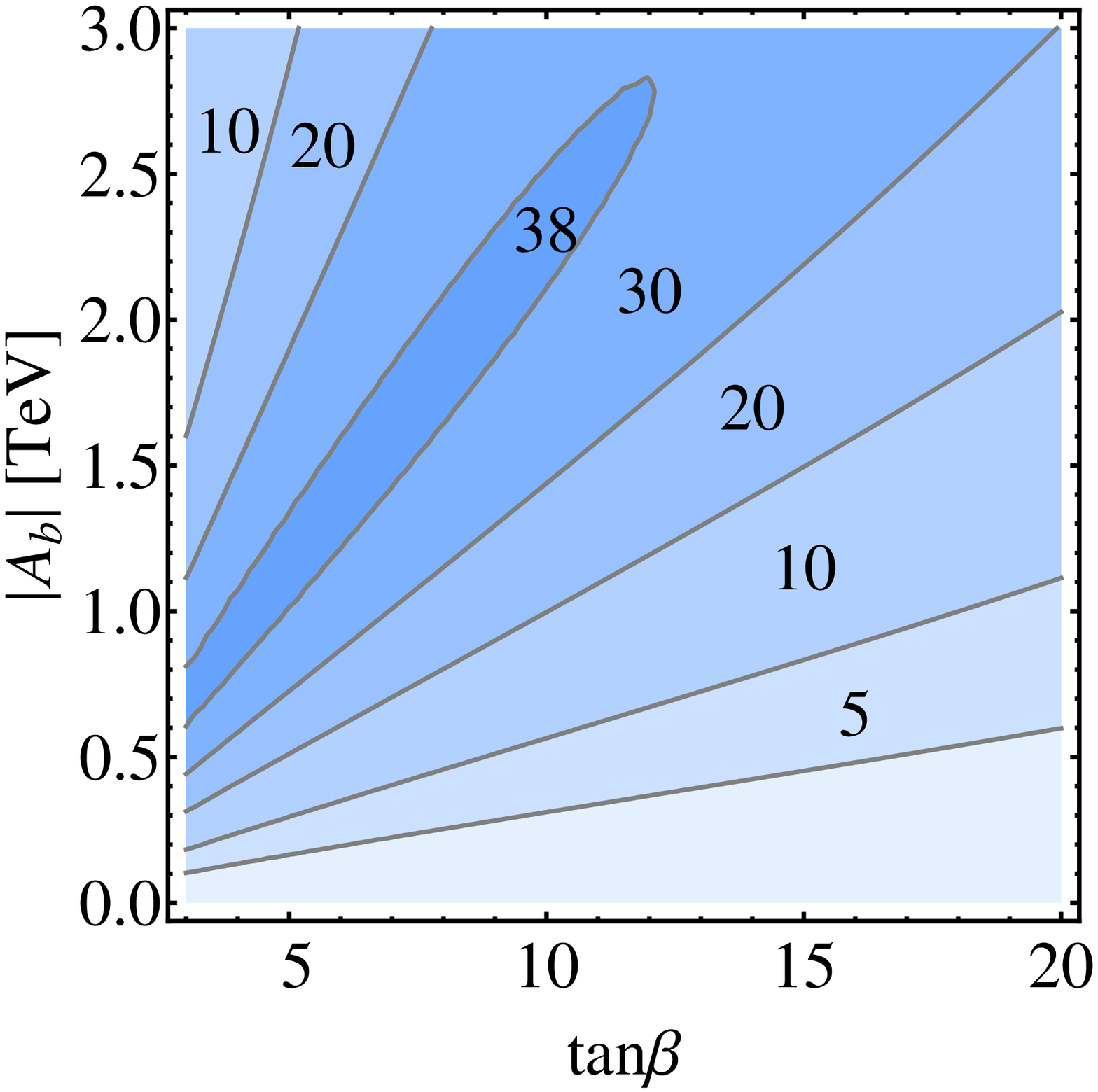}
\caption{
         Contour lines of the CP asymmetry  ${\mathcal A}(t\ell_1\ell_2)$, 
         Eq.~(\ref{eq:Toddasym}), in percent for sbottom decay  
         $\tilde b_1 \to t \tilde\chi^-_1$,  followed by the subsequent two-body decay
         chain  $\tilde\chi^-_1 \to \ell_1^- \,\tilde\nu_\ell^{\ast}$,
         and $t\to b\,\nu_\ell\,\ell_2^+ $, in the plane of 
         the soft breaking parameters $M_{\tilde Q}$, $M_{\tilde D}$ (left),
         and in the  $\tan\beta$--$|A_b|$ plane (right).
         The SUSY parameters are given in Table~\ref{tab:ReferenceScenario}.
}
\label{fig:asym_MQD_Abeta}
\end{figure}

\medskip

The size of the asymmetry ${\mathcal A}$, Eq.~(\ref{eq:Adependence2}),
strongly depends on the sbottom mixing, see the mixing matrix
Eq.~(\ref{eq:mm}), in Appendix~\ref{sec:SbottomMixing}.
The mixing is determined by the soft-breaking parameters 
$M_{\tilde Q}$, $M_{\tilde D}$, on the diagonal, and
$m_b (A_b-\mu^\ast\tan\beta)$ on the off-diagonal entries of the
sbottom mixing matrix.
In Figure~\ref{fig:asym_MQD_Abeta} (left), we see that the  
asymmetry is maximal in the region of the level crossing
$M_{\tilde Q}\approx M_{\tilde D}$, where also 
${\rm Im}\{ l^{\tilde b}_{11} (k^{\tilde b}_{11})^\ast \}$
is maximal, see  Eq.~(\ref{expansion}).
However, although the imaginary part quickly drops for increasing 
$M_{\tilde Q}> M_{\tilde D}$, the asymmetry is still sizeable
in that region, since the couplings still fulfill
$ |l^{\tilde b}_{11}| \approx |k^{\tilde b}_{11}|$.

\medskip

In Figure~\ref{fig:asym_MQD_Abeta} (right), we show in the 
$\tan\beta$--$|A_b|$ plane contour lines of the 
asymmetry, which peaks at $|A_b| \approx |\mu|\tan\beta$,
where $ |l^{\tilde b}_{11}| \approx |k^{\tilde b}_{11}|$.
Whereas the imaginary part of the couplings,
${\rm Im}\{ l^{\tilde b}_{11} (k^{\tilde b}_{11})^\ast \}$, 
steadily grows with increasing $|A_b| > |\mu|\tan\beta$,
see Eq.~(\ref{expansion}), the asymmetry is reduced in that region.
This is since the  sbottom width  $\Gamma(\tilde b_1 \to t \tilde\chi^\pm_1)$ 
increases, which enters in the  denominator of the asymmetry, 
see Eq.~(\ref{eq:Toddasym}).
Thus from  Figure~\ref{fig:asym_MQD_Abeta}, we can 
observe that the contribution from the CP-even denominator of $\eta$, 
Eq.~(\ref{eq:couplingfunct}),
to the asymmetry has an important impact on its parameter dependence.

\begin{figure}[t]
\centering
\includegraphics[clip,width=0.410\textwidth]{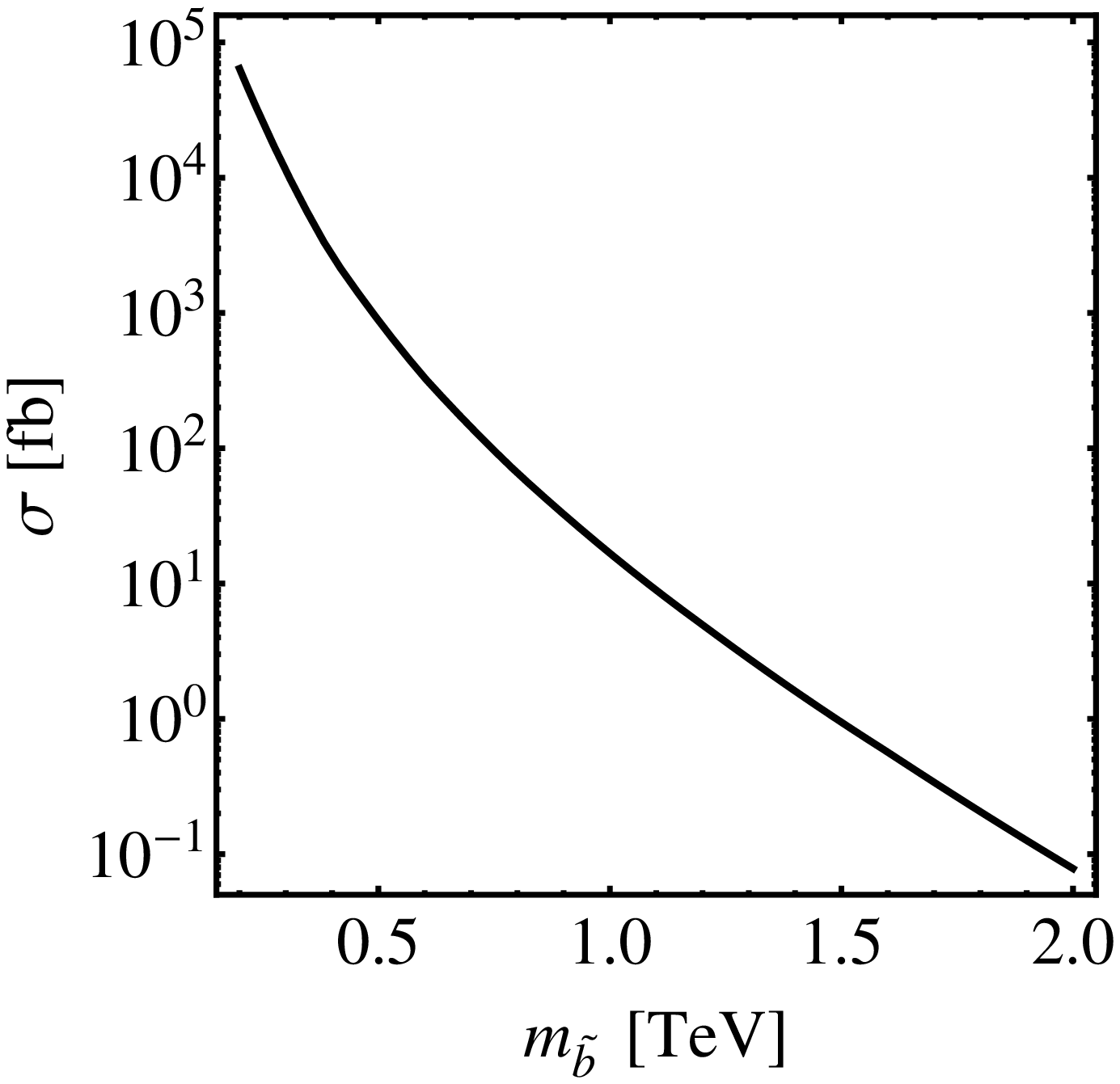}
\includegraphics[clip,width=0.580\textwidth]{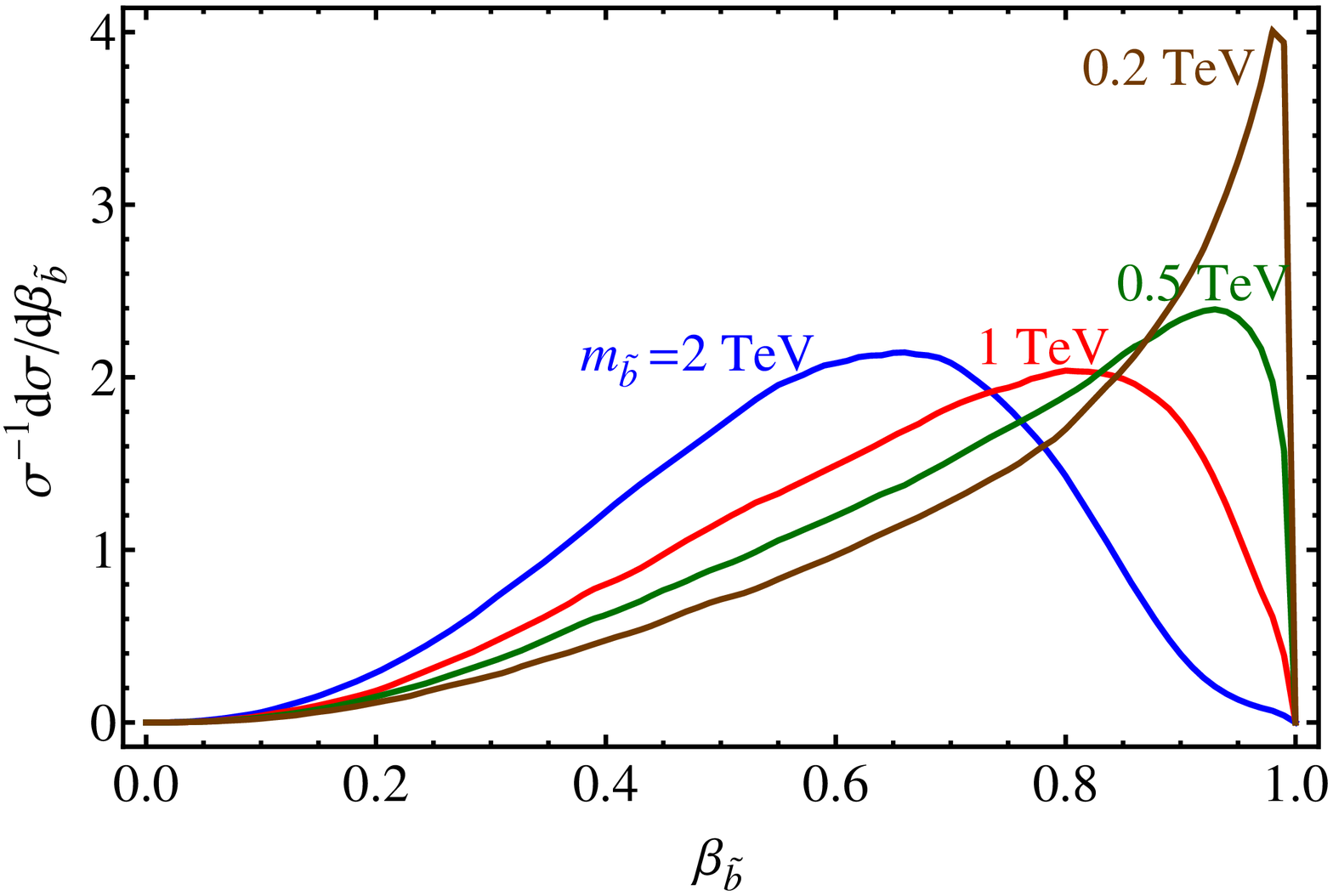}
\caption{Total sbottom pair production cross section \(\sigma(pp\to\tilde b\tilde b^*)\)
         at the LHC as a function of the sbottom mass (left). Normalized sbottom pair 
         production distribution  
         \(\frac{1}{\sigma}\frac{d\sigma}{d\beta_{\tilde b}}\) with respect to the 
         sbottom boost factor \(\beta_{\tilde b}\) (right). The leading order
         cross sections have been calculated at $\sqrt{s_{pp}}=14$~TeV using 
         MadGraph~\cite{Stelzer:1994ta}.}
\label{fig:StopPairProduction}
\end{figure}

%
\section{CP asymmetries at the LHC}
\label{sec:AsymmetryLHC}

The production of sbottom pairs at the LHC~\cite{NLO}
\begin{equation}
p + p  \to   \tilde b_m + \tilde b_m^\ast, \quad m =1,2,
\label{eq:sbottomProd}
\end{equation}
dominantly proceeds via gluon fusion.
As a result, the leading order cross section $\sigma(pp\to\tilde b_m\tilde b_m^\ast)$ 
is independent of any other SUSY model parameters than the sbottom mass.
The strong dependence can be seen in Figure~\ref{fig:StopPairProduction}~(left),
where the cross section drops by six orders of magnitude with an increase of the 
squark mass from $0.2$~TeV to $2$~TeV. 
%
%
The produced sbottoms have a distinct distribution in their boost 
\begin{equation}
\beta_{\tilde b}=\frac{|\mathbf{p}_{\tilde b}|}{E_{\tilde b}},
\label{eq:sbottomBoost}
\end{equation}
along the direction of their momenta. In  Figure~\ref{fig:StopPairProduction}~(right),
we show the normalized boost distribution for several values for the sbottom mass. 
Typically, light sbottoms are highly boosted in the laboratory frame.
In our scenario (Table~\ref{tab:ReferenceScenario}), the lightest sbottom has 
a mass of $m_{\tilde b_1}=400$~GeV and its boost distribution peaks at 
$\beta_{\tilde b_1}\approx 0.95$.

\medskip

\begin{figure}[t]
\centering
\includegraphics[clip,width=0.580\textwidth]{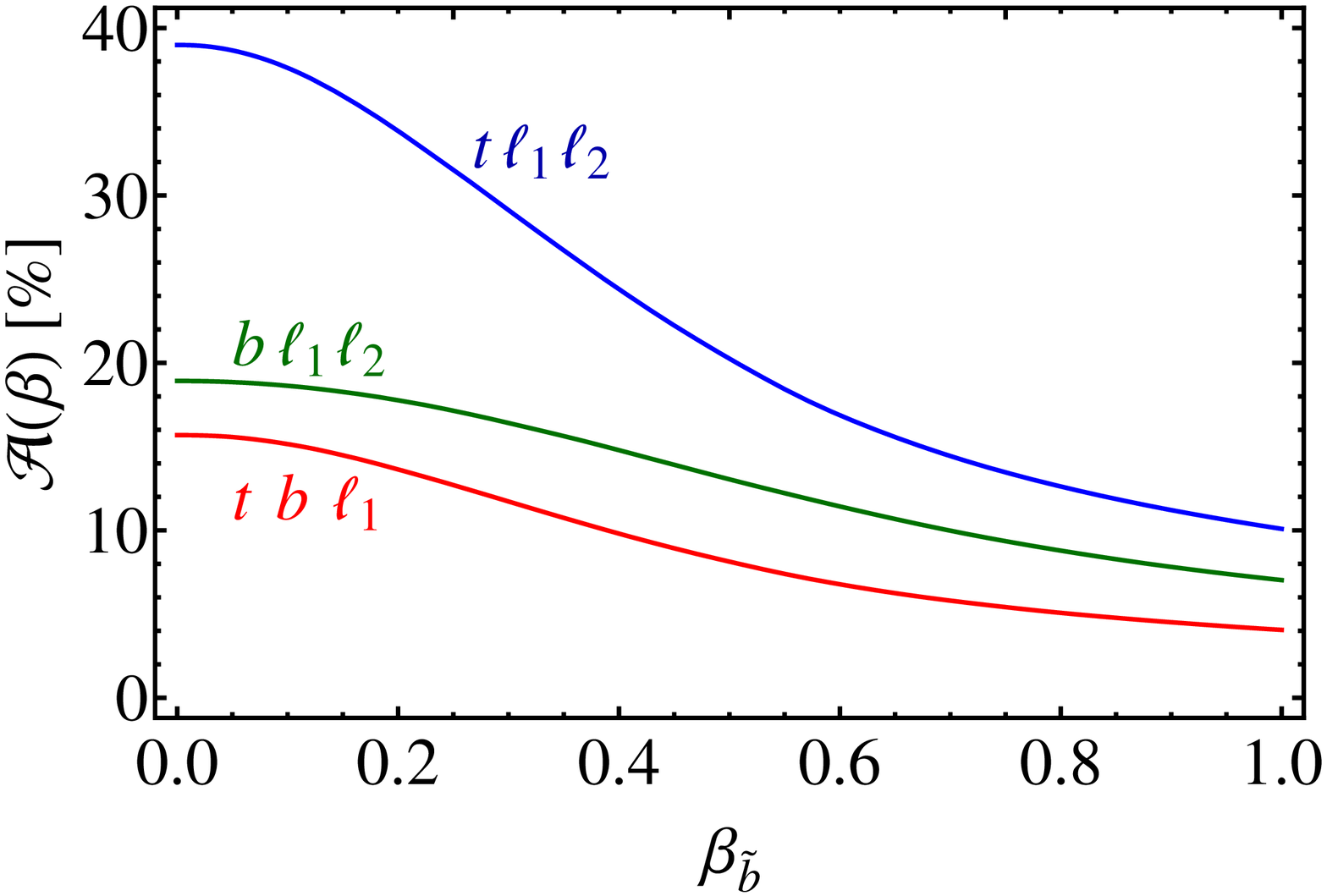}
\includegraphics[clip,width=0.410\textwidth]{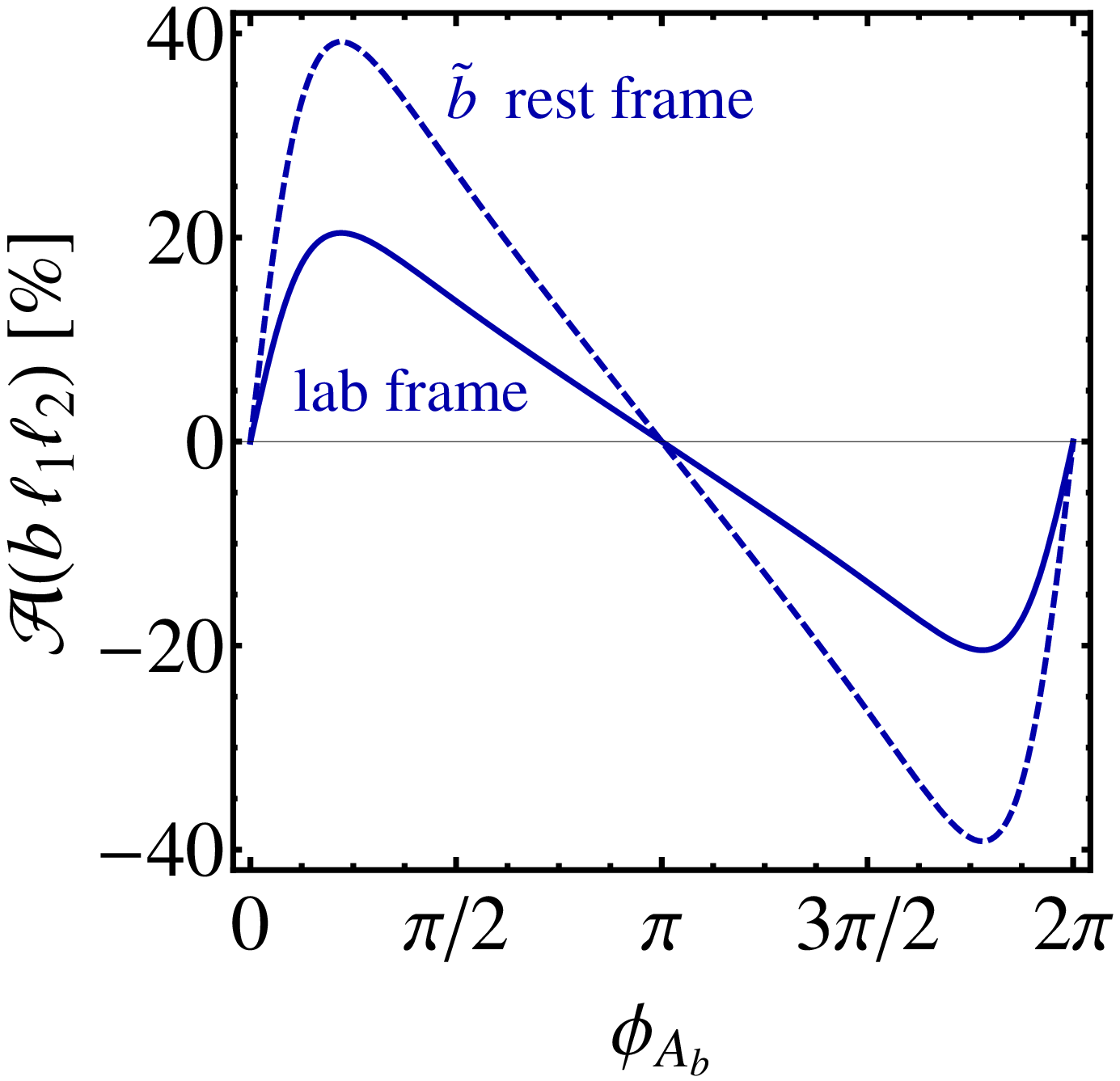}
\caption{Triple product asymmetries  
         ${\mathcal A}(t\ell_1\ell_2)$, ${\mathcal A}(b\ell_1\ell_2)$,   
         and ${\mathcal A}(t b \ell_1)$, see Eq.~(\ref{eq:Toddasym}),   
         for sbottom decay, ($\tilde b_1 \to t \tilde\chi^-_1$, 
         followed by $\tilde\chi^-_1 \to \ell_1^- \,\tilde\nu_\ell^{\ast}$,
         and $t\to b\nu_\ell\ell_2^+ $, see Figure~\ref{Fig:decaySbottom}),
         as a function of the sbottom boost 
         $\beta_{\tilde b_1}$ (left). 
         Asymmetry ${\mathcal A}(t\ell_1\ell_2)$ in the sbottom rest frame and the 
         laboratory frame (at the LHC with $\sqrt s =14$~TeV) as a function of 
         $\phi_{A_b}$ (right).
         The SUSY parameters are given in Table~\ref{tab:ReferenceScenario}.}
\label{fig:BoostedAsymmetry}
\end{figure}

The asymmetries 
which are based on epsilon products  ${\mathcal E}$,
see Eqs.~(\ref{eq:epsilon}) and (\ref{eq:Toddasym}),  are independent of the 
sbottom boost, since they are by construction Lorentz invariant. In contrast,
the triple product asymmetries are not boost invariant.
In Figure~\ref{fig:BoostedAsymmetry}~(left), we show their boost dependence
for the three possible triple product combinations 
${\mathcal T}= (t\ell_1\ell_2)$, $(b\ell_1\ell_2)$,  and $(tb\ell_1)$. 
The asymmetries are calculated for 
 $pp\to\tilde b_1\tilde b_1^\ast $ production at the LHC, with the subsequent
decays $\tilde b_1 \to t \tilde\chi^-_1$  and $\tilde\chi^-_1\to \ell_1 \tilde \nu_\ell$.
In the sbottom rest frame, $\beta_{\tilde b}=0$, they coincide with the corresponding  
epsilon product asymmetries.
Note that the size of the asymmetries strongly depends
on the choice of momenta, and largest values are obtained for the optimal
triple product ${\mathcal T}= (t\ell_1\ell_2)$, as given in Eq.~(\ref{tripleterm}).

%

\subsection{Triple product asymmetries in the laboratory frame}
\label{sec:AsymmetryLab}
%
The size of the triple product asymmetry in the laboratory (lab) frame of the LHC, 
is obtained by folding the boost dependent asymmetry ${\mathcal A}(\beta_{\tilde b})$
with the normalized sbottom boost distribution~\cite{Deppisch:2009nj},
\begin{equation}
	{\mathcal A}^{\rm lab} =
	\frac{1}{\sigma}
	\int_{0}^{1} 
	\frac{d\sigma}{d\beta_{\tilde b}}
	~{\mathcal A}(\beta_{\tilde b})
	~d\beta_{\tilde b},
\label{eq:boostasymmetry}
\end{equation}
of the production cross section  $\sigma=\sigma(pp\to\tilde b_m\tilde b_m^\ast)$.
The folded asymmetry ${\mathcal A}^{\rm lab}$ is almost reduced by half,
compared to ${\mathcal A}(\beta_{\tilde b}=0)$.
This can be seen in Figure~\ref{fig:BoostedAsymmetry}~(right), where we show
both asymmetries as a function of the CP phase $\phi_{A_b}$. 
Thus the sbottom boost effectively reduces the asymmetries, but does not change their 
shape with respect to the phase dependence.
\medskip

In the following, we will quantify the expected measurability of the 
the folded triple product asymmetry  
${\mathcal A}^{\rm lab} (t\ell_1\ell_2)$, Eq.~(\ref{eq:boostasymmetry}), 
in the lab frame. 
Since a measurement of the asymmetries also depends on the production cross section,
the sbottom boost distribution, and the size of the corresponding
branching ratios for sbottoms and charginos,
we will present the upper bounds on the expected luminosity,
\begin{eqnarray}
{\mathcal L} &=& \frac{1}{\sigma} 
          \left( \frac{1}{{\mathcal A^2}} -1 \right).
\label{eq:lumi}
\end{eqnarray}
As defined in Appendix~\ref{sec:Significances}, ${\mathcal L}$ is the minimal required
luminosity to observe a signal at $95\%$ CL above statistical fluctuations.
In  Eq.~(\ref{eq:lumi}), the combined cross section of sbottom production and decay
is denoted by $\sigma$.

\medskip

\begin{figure}[t!]
\centering
\includegraphics[clip,width=0.495\textwidth]{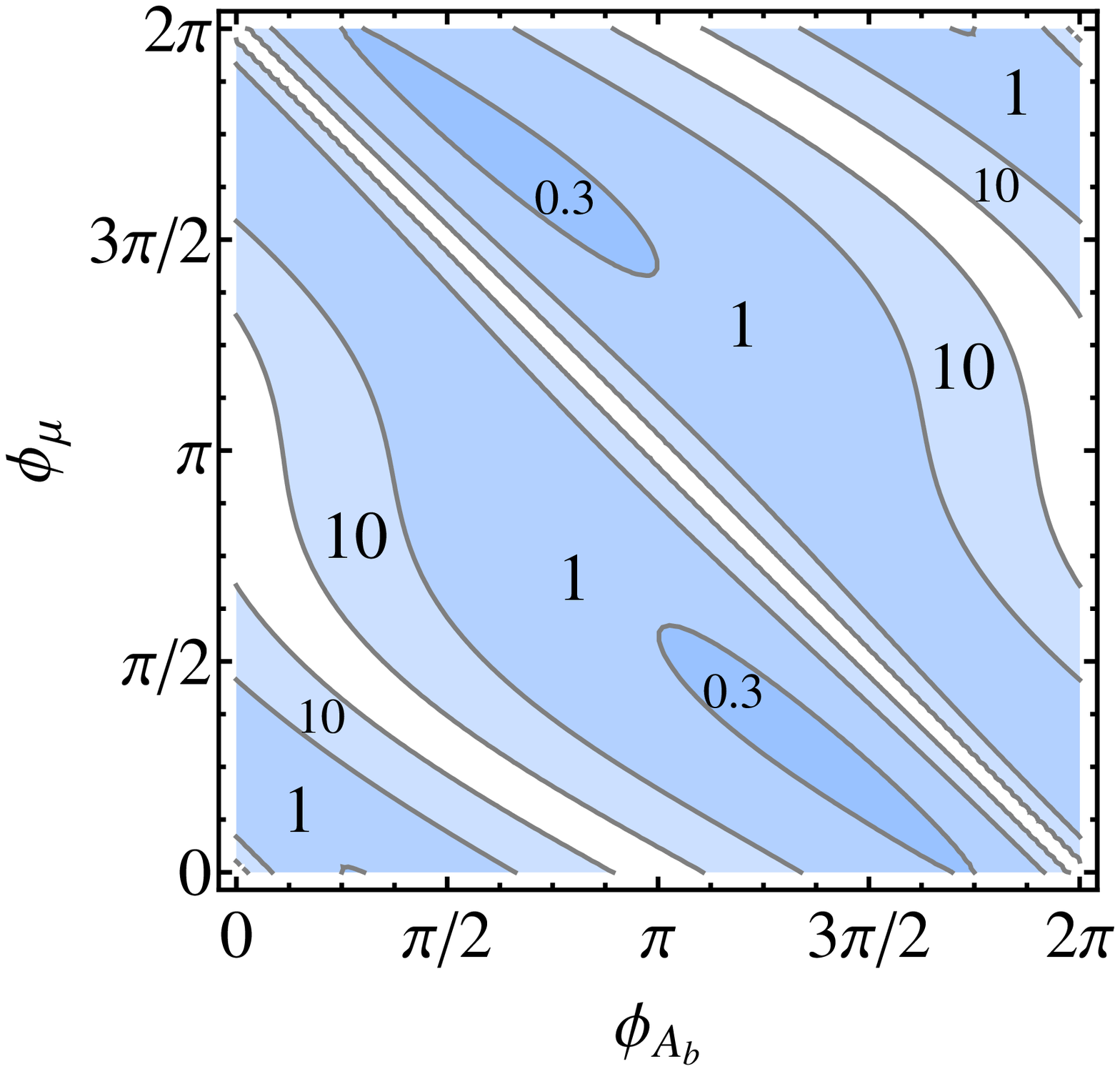}
\includegraphics[clip,width=0.495\textwidth]{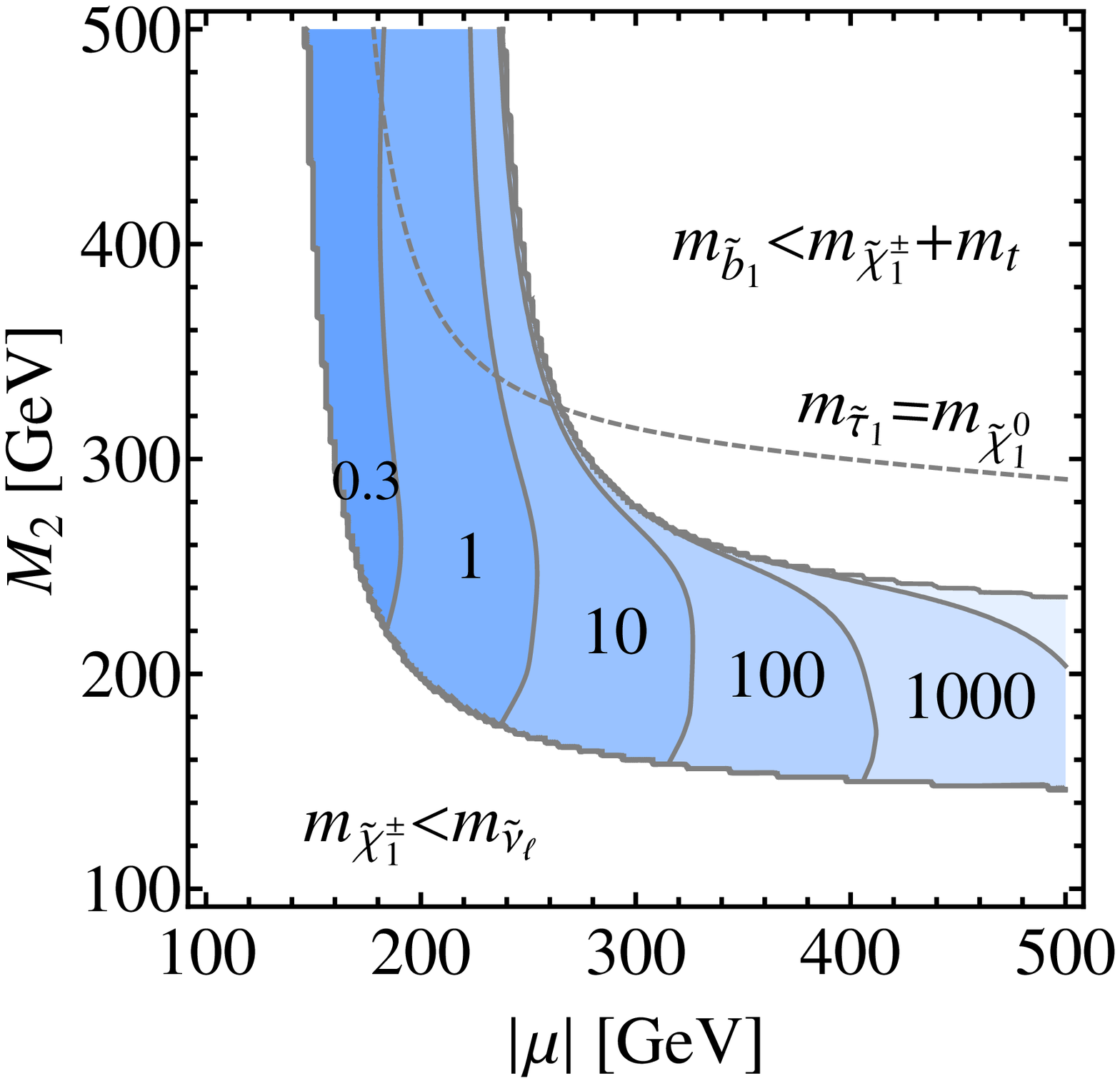}
\includegraphics[clip,width=0.495\textwidth]{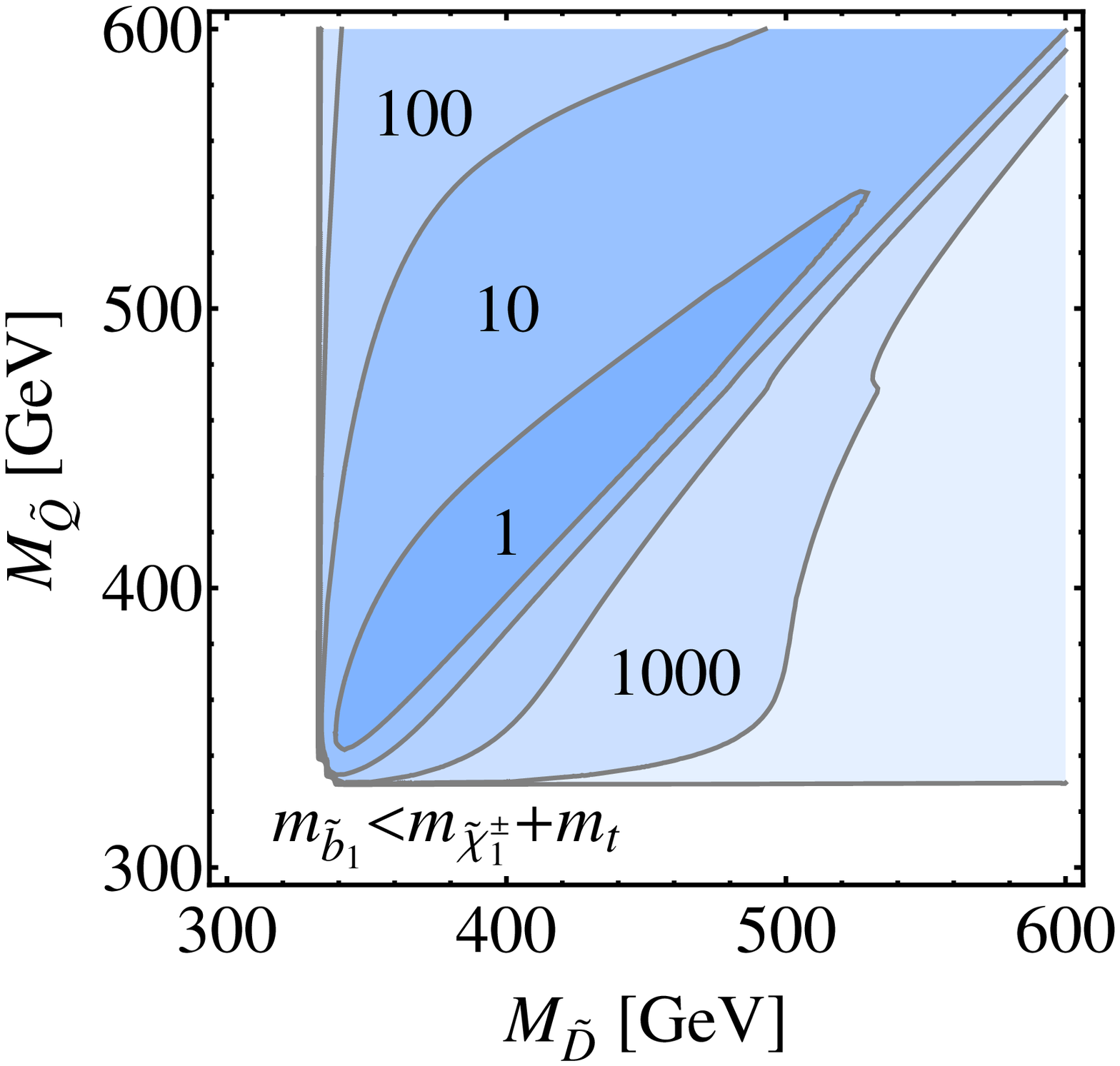}
\includegraphics[clip,width=0.475\textwidth]{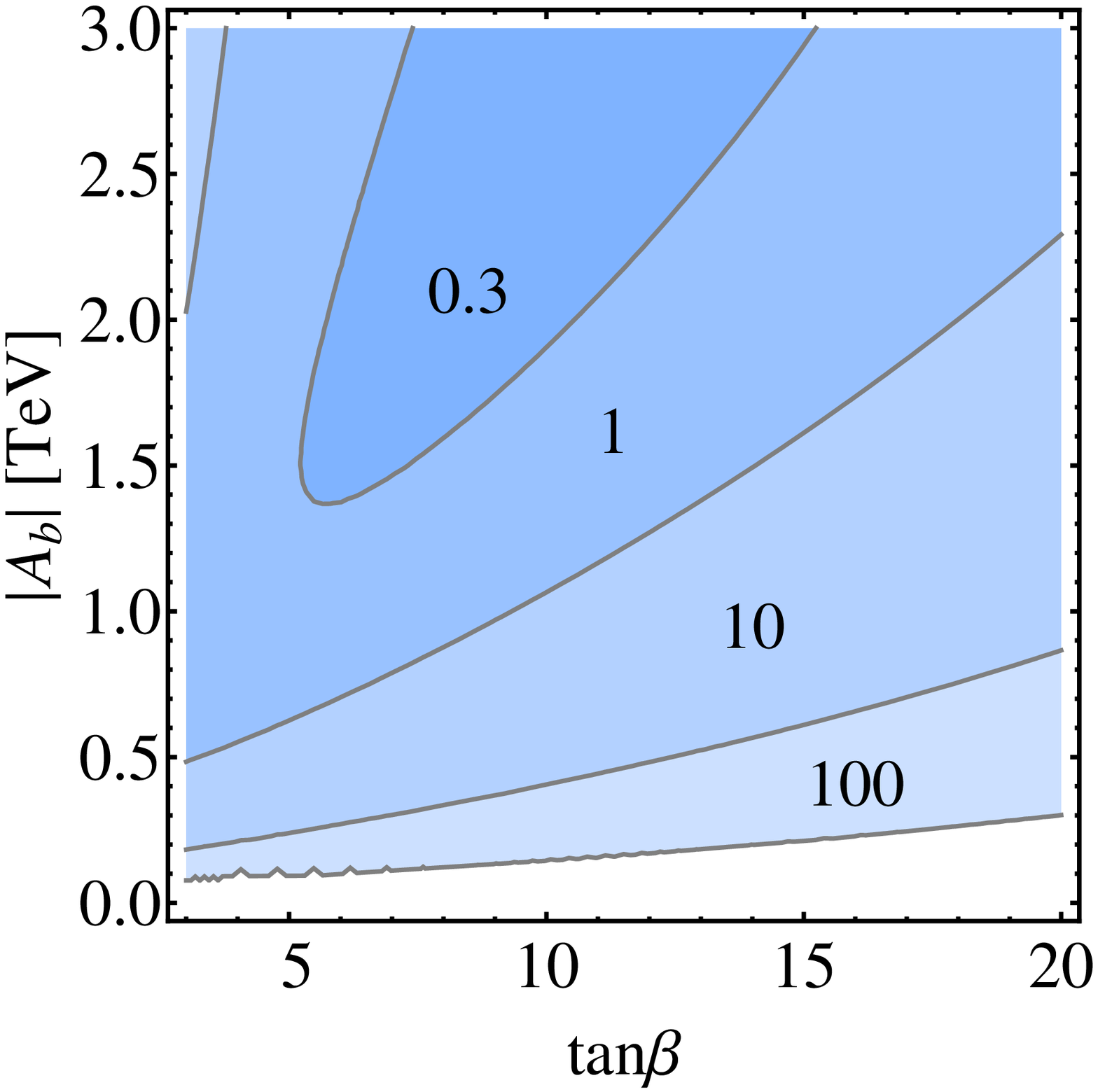}
\caption{
         Contour lines of the minimal required luminosity  
         ${\mathcal L}$, Eq.~(\ref{eq:lumi}), in units of fb$^{-1}$ to observe
         the triple product asymmetry 
         $\mathcal{A}^{\rm lab}(t\ell_1\ell_2)$, Eq.~(\ref{eq:boostasymmetry}), 
         at $1\sigma$ above fluctuations 
         at the LHC, with $\sqrt{s_{pp}}=14$~TeV,
         for sbottom production and decay,
         $\tilde b_1 \to t \tilde\chi^-_1$, followed by 
         $\tilde\chi^-_1 \to \ell_1^- \,\tilde\nu_\ell^{\ast}$,
         and $t\to b\,\nu_\ell\,\ell_2^+ $.
         The SUSY parameters are given in Table~\ref{tab:ReferenceScenario}.
}
\label{fig:lumis}
\end{figure}

In Figure~\ref{fig:lumis}, we show contour lines of the minimal required
luminosity ${\mathcal L}$ in different planes of the 
MSSM parameter space for the benchmark scenario as 
given in Table~\ref{tab:ReferenceScenario}.
The typical sizes of the sbottom and chargino branching ratios
${\rm BR}(\tilde b_1 \to t \tilde\chi^\pm_1)$, and
${\rm BR}(\tilde\chi^\pm_i \to e_1^\pm \tilde\nu_e)$,
are of the order of $15\%$ to $35\%$.
As can be seen, a large part of the parameter space can be
probed with  ${\mathcal L} = 10~{\rm fb}^{-1}$.

\medskip

Note however that the presented values must be regarded as absolute lower bounds
on the minimal required luminosities, only.
The definition of ${\mathcal L}$, as in Eq.~(\ref{eq:lumi}), is purely based on the
 theoretical signal rate and its asymmetry, since  detector efficiency
 effects and contributions from CP-even backgrounds are not included.
However, the minimal luminosity requirement  can be used to exclude those  MSSM parameter 
regions which cannot be probed at the LHC. 
Clearly, in order to give realistic values of the statistical significances and required 
luminosities, a detailed experimental study is necessary, which is however beyond 
the scope of the present work.
In particular, the feasibility of the event reconstruction has to be discussed. 
First ideas to  attempt the reconstruction of a stop decay chain 
have recently been reported~\cite{MoortgatPick:2009jy}.


\section{Summary and conclusions}
\label{sec:Conclusion}

\medskip

We have analyzed CP observables in the two-body decays of a light sbottom
\begin{eqnarray}
	\tilde b_1 \to t + \tilde\chi^-_1.
\end{eqnarray}
The CP-sensitive parts appear only in the top chargino spin-spin correlations, 
which can be probed by the subsequent decays 
\begin{eqnarray}
\tilde\chi^-_i  &\to  & \ell_1^- + \tilde\nu_\ell^\ast, \quad 
                        \tilde\nu_\ell^\ast \to \tilde\chi^0_1 + \bar\nu_\ell,
                         \quad  \ell= e,\mu,\nonumber\\[2mm]
	                t  &\to& b + W, \;  \quad W \to  \nu_\ell + \ell_2.
\end{eqnarray}
Due to angular momentum conservation, the decay distributions of the final state momenta 
are correlated to each other. Asymmetries of triple products of three spatial momenta, 
as well as epsilon products of four space-time momenta are ideal tools
to probe the CP-sensitive  spin-spin correlations.
The CP observables we have proposed are sensitive to the CP phases of the 
trilinear coupling parameter $A_b$, 
and the higgsino mass parameter $\mu$, which might be present in 
the sbottom and chargino sector of the MSSM. 

\medskip

We have analyzed the asymmetries and event rates in a general MSSM framework.
For maximal sbottom mixing,  the asymmetries reach up to $40\%$.
The class of asymmetries which are based on  triple products 
are not Lorentz invariant, and thus are reduced by a 
factor of about three when evaluated in the laboratory frame at the LHC. 
Luminosities of at least $10 \textrm{ fb}^{-1}$ are required 
to observe a CP signal of $1\sigma$ above statistical fluctuations at the LHC.

\medskip

Clearly, the measurability of the asymmetries and thus of the CP phases can only
be addressed properly in a detailed experimental analysis, which should take into 
account background processes, detector simulations and event reconstruction efficiencies.
 We want to stress the need for such a thorough analysis, to explore the potential 
of the LHC to probe SUSY CP violation.

\medskip

Interrelations between the T-violating electric dipole moments (EDMs) 
and the possible SUSY CP phases underline the need to determine CP observables
outside the low energy sector, in particular by measurements at the LHC.
Since the proposed CP asymmetries at colliders 
depend in a different way on the SUSY parameters than the EDMs,
they would be the ideal tools for such independent measurements.
Depending on the experimental results, the EDM  bounds could be either verified,  
or the CP-violating sectors of the underlying  SUSY models have to be modified.  

\subsection*{Acknowledgments}
We thank F.~von~der~Pahlen for very helpful comments and discussions. 
This work was supported by MICINN project FPA.2006-05294.
The authors thank the Aspen Center for Physics for hospitality. \\

\medskip

\begin{appendix}
\noindent{\Large\bf Appendix}
\setcounter{equation}{0}
\renewcommand{\thesubsection}{\Alph{section}.\arabic{subsection}}
\renewcommand{\theequation}{\Alph{section}.\arabic{equation}}
\setcounter{equation}{0}

\section{Sbottom mixing}
\label{sec:SbottomMixing}

The masses and couplings of the sbottoms follow from their mass matrix~\cite{Bartl:2006hh}
\begin{eqnarray}
	{\mathcal{L}}_M^{\tilde b} &=&
	-(\tilde b_L^{\ast},\, \tilde b_R^{\ast})
	\left(\begin{array}{ccc}
		m_{\tilde b_{L}}^2 & 
		e^{-i\phi_{\tilde b}}m_b |\Lambda_{\tilde b}|\\[5mm]
		e^{ i\phi_{\tilde b}}m_b |\Lambda_{\tilde b}| & 
		m_{\tilde b_R}^2
	\end{array}\right)
	\left(\begin{array}{ccc}
		\tilde b_L\\[5mm]
		\tilde b_R 
	\end{array}\right),
\label{eq:mm}
\end{eqnarray}
with 
\begin{eqnarray}
	m_{\tilde b_L}^2 & = & 
	M_{\tilde Q}^2+
	\left(-\frac{1}{2}+\frac{1}{3}\sin^2\theta_w \right)
	m_Z^2\cos(2\beta)  +m_b^2, 
\label{eq:ml} \\[2mm]
	m_{\tilde b_R}^2 & = & 
	M_{\tilde D}^2-\frac{1}{3}m_Z^2\sin^2\theta_w\cos(2\beta) + m_b^2,
\label{eq:mr}
\end{eqnarray}
with the soft SUSY-breaking parameters $M_{\tilde Q}$, $M_{\tilde D}$, the ratio 
$\tan\beta=v_2/v_1$ of the vacuum expectation values of the two neutral 
Higgs fields, the weak mixing angle $\theta_w$, the mass $m_Z$ of the $Z$ boson, 
and the mass $m_b$ of the bottom quark. The CP phase of the sbottom sector is
\begin{eqnarray}\label{eq:phib}
	   \phi_{\tilde b}& = & \arg\lbrack \Lambda_{\tilde b}\rbrack,\\ [2mm]
	\Lambda_{\tilde b}& = & A_b-\mu^\ast\tan\beta,
\label{eq:mlr}
\end{eqnarray}
with the complex trilinear scalar coupling parameter $A_b$, and the higgsino mass 
parameter $\mu$. 
Note that for $|A_b|\gg |\mu| \tan\beta$ we have $\phi_{\tilde b}\approx\phi_{A_b}$.
The sbottom mass eigenstates 
\begin{eqnarray}
	\left(\begin{array}{ccc}
		\tilde b_1 \\[1mm]
		\tilde b_2 
	\end{array}\right)
	&=&
	\mathcal{R}^{\tilde b}
	\left(\begin{array}{ccc}
		\tilde b_L \\[1mm]
		\tilde b_R 
	\end{array}\right),
\label{eq:sbottommasseigen}
\end{eqnarray}
are given by the diagonalization matrix~\cite{Bartl:2006hh}
\begin{equation}
	\mathcal{R}^{\tilde b} = 
	\left( \begin{array}{ccc}
		e^{i\phi_{\tilde b}}\cos\theta_{\tilde b} &
		\sin\theta_{\tilde b}\\[5mm]
	       -\sin\theta_{\tilde b} &
		e^{-i\phi_{\tilde b}}\cos\theta_{\tilde b}
	\end{array}\right),
\label{eq:rsbottom}
\end{equation}
with the sbottom mixing angle $\theta_{\tilde b}$
\begin{equation}
	\cos\theta_{\tilde b}=
	\frac{-m_b |\Lambda_{\tilde b}|}{\sqrt{m_b^2 |\Lambda_{\tilde b}|^2+
	(m_{\tilde b_1}^2-m_{\tilde b_{L}}^2)^2}},\quad
	\sin\theta_{\tilde b}=\frac{m_{\tilde b_{L}}^2-m_{\tilde b_1}^2}
	{\sqrt{m_b^2 |\Lambda_{\tilde b}|^2+(m_{\tilde b_1}^2 
	-m_{\tilde b_{L}}^2)^2}}.
\label{eq:thsbottom}
\end{equation}
The mass eigenvalues are
\begin{equation}
	m_{\tilde b_{\,1,2}}^2 = 
	\frac{1}{2}\left[(m_{\tilde b_{L}}^2+m_{\tilde b_R}^2)\mp 
	\sqrt{(m_{\tilde b_{L}}^2 - m_{\tilde b_{R}}^2)^2
	+4m_b^2 |\Lambda_{\tilde b}|^2}\;\right].
\label{eq:m12}
\end{equation}

\section{Chargino mixing}
\label{sec:CharginoMixing}

The complex mass matrix of the charginos is~\cite{mssm}  
\begin{eqnarray} 
M_{\chi^\pm}= 
\left(\begin{array}{cc} 
      M_2                   &  m_W\sqrt{2}\sin{\beta}\\ 
      m_W\sqrt{2}\cos{\beta}& \mu 
\end{array}\right). 
\label{eq:charmassmatrix}
\end{eqnarray} 
with the $SU(2)$ gaugino mass parameter $M_2$, and the mass of the $W$ boson $m_W$.
We optain the chargino masses and the couplings
by diagonalizing the chargino matrix~\cite{mssm},
\begin{equation} 
U^{\ast} M_{\chi^\pm} V^{\dagger} = 
\textsl{diag}(m_{\chi^\pm_{1}},m_{\chi^\pm_{2}}),
\label{diag}   
\end{equation} 
with the two independent, unitary diagonalization matrices $U$ and $V$.

\section{Lagrangians and complex couplings}
\label{Lagrangians and couplings}

The interaction Lagrangian for sbottom decay
$\tilde b_m \to t \tilde\chi_i^- $, $\tilde b_m^\ast \to \bar t \tilde\chi_i^+ $,
is given by~\cite{Bartl:2006hh}
\begin{eqnarray}
{\scr L}_{t \tilde b\tilde\chi^+}
= g\,\bar t\,(l_{mi}^{\tilde b}\,P_R + k_{mi}^{\tilde b}\,P_L)
            \,\tilde\chi_i^+\,\tilde b_m + {\rm h.c.} ,
\end{eqnarray}
with $P_{L, R}=(1\mp \gamma_5)/2$, and the weak coupling constant
$g=e/\sin\theta_w$, $e>0$. The couplings are
\begin{equation}
l_{mi}^{\tilde b} = -{\mathcal R}^{\tilde b \ast}_{m1}\, U_{i1}+
Y_b\, {\mathcal R}^{\tilde b\ast}_{m2}\, U_{i2}, \qquad
k_{mi}^{\tilde b} =Y_t\,{\mathcal R}^{\tilde b \ast}_{m1}\,V_{i2}^{\ast},
\label{eq:lksbottom}
\end{equation}
with the sbottom diagonalization matrix ${\mathcal R}^{\tilde b}$,  Eq.~(\ref{eq:rsbottom}),
the chargino diagonalization matrices $U,V$,  Eq.~(\ref{diag}),
and the Yukawa couplings
\begin{equation}
      Y_t  = \frac{m_t}{\sqrt{2}\,m_W \sin\beta}, \quad 
      Y_b  = \frac{m_b}{\sqrt{2}\,m_W \cos\beta}.
\label{eq:ytb}
\end{equation}
The interaction Lagrangians for chargino decay
$\tilde\chi_i^\pm \to \ell^\pm \tilde\nu_\ell^{(\ast)} $, or
$\tilde\chi_i^\pm \to \tilde\ell_L^\pm \nu_\ell$, are~\cite{mssm}
\begin{eqnarray}
{\scr L}_{\ell \tilde\nu\tilde\chi^+_i} &=&
	- g U_{i1}^{*} \bar{\tilde\chi}^+_i P_{L} \nu \tilde\ell_L^*
	- g V_{i1}^{*} \bar{\tilde\chi}_i^{+C} P_L \ell 
				 \tilde{\nu}^{*}+\mbox{h.c.},\quad \ell=e,\mu.
\label{eq:slechie}
\end{eqnarray}

\section{Kinematics and phase space}
\label{Phase space}

For the sbottom decay $\tilde b_m \to t\tilde\chi_i^\pm$,
we choose a coordinate frame in the laboratory~(lab) frame
such that the momentum of the sbottom $\tilde b_m$
points in the $z$-direction 
\begin{eqnarray}
  p_{\tilde b}^{\mu} &=& (E_{\tilde b},\, 0,\, 0,\, |{\mathbf p}_{\tilde b}| ), \\[2mm]
  p_t^{\mu} &=& 
   (E_t,\, |{\mathbf p}_t|\sin\theta_t,\, 0,\, |{\mathbf p}_t|\cos\theta_t). 
\label{eq:momenta1}
\end{eqnarray}
The decay angle 
$\theta_t =\varangle ({\mathbf p}_{\tilde b},{\mathbf p}_t)$
of the top quark is constrained by
$\sin\theta^{\rm max}_t= |{\mathbf p}_{\tilde b}^\prime | /
                         |{\mathbf p}_{\tilde b}|$
for $|{\mathbf p}_{\tilde b}|>|{\mathbf p}_{\tilde b}^\prime|=
\lambda^{\frac{1}{2}}(m^2_{\tilde b},m^2_t,m^2_{\chi^\pm_i})/2m_t$,
with the triangle function $\lambda(a,b,c)=a^2+b^2+c^2-2(a b + a c + b c)$.
In this case there are two solutions~\cite{Kittel:2004rp,Byckling} 
\begin{eqnarray}
| {\mathbf p}^{\pm}_t|&=& \frac{
(m^2_{\tilde b}+m^2_t-m^2_{\chi^\pm_i}) |{\mathbf p}_{\tilde b}|\cos\theta_t\pm
E_{\tilde b}\sqrt{\lambda(m^2_{\tilde b},m^2_t,m^2_{\chi^\pm_i})-
         4|{\mathbf p}_{\tilde b}|^2~m^2_t~\sin^2\theta_t}}
        {2|{\mathbf p}_{\tilde b}|^2 \sin^2\theta_t
         +2 m^2_{\tilde b}}.\nonumber \\
&&
\label{maxpt}
\end{eqnarray}
For $|{\mathbf p}_{\tilde b}|<|{\mathbf p}_{\tilde b}^\prime|$
the angle $\theta_t$ is unbounded, and only the physical solution 
$ |{\mathbf p}^+_t|$ is left.

The momenta of the subsequent decays of the chargino
$\tilde\chi_i^\pm  \to \ell_1 \tilde\nu_\ell^{(\ast)}$,
Eq.~(\ref{eq:decayChi}),
and those of the top quark
$t \to b W$, $W \to \ell_2 \nu_\ell$, Eq.~(\ref{eq:decayTop}), 
can be parametrized by
\begin{eqnarray}
  p_b^{\mu} &=& E_b(1,\, \sin \theta_b \cos \phi_b,
                      \, \sin \theta_b \sin \phi_b,
                      \, \cos \theta_b), \\[2mm]
  p_{\ell_1}^{\mu} &=& E_{\ell_1}(1,\, \sin \theta_1 \cos \phi_1,
                      \, \sin \theta_1 \sin \phi_1,
                      \, \cos \theta_1), \\[2mm]
  p_{\ell_2}^{\mu} &=& E_{\ell_2}(1,\, \sin \theta_2 \cos \phi_2,
                      \, \sin \theta_2 \sin \phi_2,
                      \, \cos \theta_2),
\label{eq:momenta2}
\end{eqnarray}
with the energies~\cite{Kittel:2004rp}
\begin{eqnarray}
E_{\ell_1} = \frac{ m^2_{\chi_i^\pm}-m^2_{\tilde\nu_\ell} }
            {2(E_{\chi_i^\pm}-|{\mathbf p}_{\chi_i^\pm}|\cos\theta_{D_1}  )},&&\quad
%
       E_b = \frac{ m^2_t-m_W^2 }
            {2(E_t-|{\mathbf p}_t|\cos\theta_{D_b}  )}, 
\\[2mm]
E_{\ell_2} = \frac{ m^2_W }
             {2(E_W-|{\mathbf p}_W|\cos\theta_{D_2}  )}. &&
\label{eq:energies}
\end{eqnarray}
The decay angles,
$\theta_{D_1} =\varangle ({\mathbf p}_{\chi_i^\pm},{\mathbf p}_{\ell_1})$,
$\theta_{D_b} =\varangle ({\mathbf p}_{t},{\mathbf p}_b)$, and
$\theta_{D_2} =\varangle ({\mathbf p}_W,{\mathbf p}_{\ell_2})$, are
\begin{eqnarray}
\cos\theta_{D_1} = 
\hat{\mathbf p}_{\chi_i^\pm}\cdot\hat{\mathbf p}_{\ell_1}, \quad
\cos\theta_{D_b} = 
\hat{\mathbf p}_{t}\cdot\hat{\mathbf p}_{b}, \quad
\cos\theta_{D_2} = 
\hat{\mathbf p}_{W}\cdot\hat{\mathbf p}_{\ell_2}, 
%
\label{eq:decayangle}
\end{eqnarray}
with the unit momentum vectors 
$\hat{\mathbf p} = {\mathbf p}/|{\mathbf p}|$,
and ${\mathbf p}_{\chi_i^\pm}={\mathbf p}_{\tilde b}-{\mathbf p}_{t}$,
 ${\mathbf p}_W= {\mathbf p}_{t}-{\mathbf p}_{b}$.

The Lorentz invariant phase-space element
for the squark decay chain, see Eqs.~(\ref{eq:decaySbottom})-(\ref{eq:decayTop}),
can be decomposed into two-body  phase-space elements~\cite{Kittel:2004rp,Byckling}
\begin{eqnarray}
 d{\rm Lips}(s_{\tilde b}\,;
p_{\nu_\ell},p_{\tilde\nu_\ell}, p_b, p_{\ell_1},p_{\ell_2})=
\frac{1}{(2\pi)^3}~
\sum_{\pm}~d{\rm Lips}(s_{\tilde b}\,;p_t,p_{\chi_i^\pm})~
\nonumber \\  \times
d s_{\chi_i^\pm}~d{\rm Lips}(s_{\chi_i^\pm};p_{\ell_1},p_{\tilde\nu_\ell})~
d s_{t}~d{\rm Lips}(s_{t};p_{b},p_{W})~
d s_{W}~d{\rm Lips}(s_{W};p_{\ell_2},p_{\nu_\ell}),
\label{eq:phasespace}
 \end{eqnarray}
where we have to sum the two solutions 
$| {\mathbf p}^{\pm}_t|$ of the top quark momentum,
see Eq.~(\ref{maxpt}),
if the decay angle $\theta_t$ is constrained.
The different factors are
 \begin{eqnarray}
d{\rm Lips}(s_{\tilde b}\,;p_t,p_{\chi_i^\pm})&=&
\frac{1}{8\pi}~
\frac{|{\mathbf p}_t|^2}{|E_t~|{\mathbf p}_{\tilde b}|\cos\theta_t-
        E_{\tilde b}~|{\mathbf  p_t}||}~\sin\theta_t~ d\theta_t,\\
d{\rm Lips}(s_{\chi_i^\pm};p_{\ell_1},p_{\tilde\nu_\ell})&=&
        \frac{1}{2(2\pi)^2}~
        \frac{|{\mathbf p}_{\ell_1}|^2}{m_{\chi_i^\pm}^2-m_{\tilde\nu_\ell}^2}
        ~d\Omega_1,\\
        d{\rm Lips}(s_{t};p_{b},p_{W})&=&
\frac{1}{2(2\pi)^2}~
        \frac{|{\mathbf p}_b|^2}{m_t^2-m_W^2}
        ~d\Omega_b,\\
        d{\rm Lips}(s_W;p_{\ell_2},p_{\nu_\ell})&=&
\frac{1}{2(2\pi)^2}~\frac{|{\mathbf p}_{\ell_2}|^2}{m_W^2}
        ~d\Omega_3,
\end{eqnarray}
with $s_j=p^2_j$ and 
$ d\Omega_j=\sin\theta_j~ d\theta_j~ d\phi_j$.
We use the narrow width approximation 
 \begin{eqnarray}
\int|\Delta(j)|^2 ~ d s_j &=& 
\frac{\pi}{m_j\Gamma_j},
\label{narrowwidth}
\end{eqnarray}
for the propagators
\begin{eqnarray}
     \Delta(j) &=& \frac{i}{s_j -m_j^2 +im_j\Gamma_j},
\label{eq:propagators}
\end{eqnarray}
which is justified for $\Gamma_j/m_j\ll1$,
which holds in our case for particle widths
$\Gamma_j\lsim {\mathcal O}(1~{\rm GeV}) $,
and masses
$m_j\approx {\mathcal O}(100~{\rm GeV}) $.
Note, however, that the naive
${\mathcal O}(\Gamma/m)$-expectation of the error can easily receive
large off-shell corrections of an order of magnitude and more,
in particular at threshold, or due to interferences
with other resonant or non-resonant processes.
For a recent discussion of these issues, see, for example, 
Ref.~\cite{narrowwidth}.

\section{Density matrix formalism}
  \label{Density matrix formalism}
%

The amplitude squared for the entire sbottom decay chain, 
Eqs.~(\ref{eq:decaySbottom})-(\ref{eq:decayTop}), has been calculated in 
Ref.~\cite{Bartl:2006hh}, by using the spin formalism of Kawasaki, Shirafuji 
and Tsai~\cite{Kawasaki:1973hf}. We calculate the amplitude squared 
in the spin-density matrix formalism~\cite{Haber:1994pe,gudi},
which allows a separation of the amplitude  squared 
into contributions from spin correlations,  spin-spin correlations and
from the unpolarized part. In that way, the CP-sensitive parts of the amplitude 
squared can be easily separated and identified, allowing
to find the epsilon product, that yields the largest CP asymmetries.

\medskip

In the  spin-density matrix formalism of Ref.~\cite{Haber:1994pe},
the amplitude squared of the sbottom decay chain, 
Eqs.~(\ref{eq:decaySbottom})-(\ref{eq:decayTop}), can be written as
\begin{eqnarray}
        |T|^2=|\Delta(t)|^2~|\Delta(\tilde\chi^-_i)|^2~|\Delta(W)|^2
 \times \nonumber \\ [3mm]
        \sum_{\lambda_i,\lambda^\prime_i,\lambda_t,\lambda^\prime_t,
              \lambda_k\lambda^\prime_k}~
        \rho_{D  }(\tilde b )_{\lambda_i\lambda^\prime_i}^{\lambda_t\lambda^\prime_t}~
        \rho_{D_1}(\tilde\chi_i^\pm)^{\lambda^\prime_i\lambda_i}~
         \rho_{D_2}(t)_{\lambda^\prime_t\lambda_t}^{\lambda^\prime_k\lambda_k}~
         \rho_{D_3}(W)_{\lambda_k\lambda^\prime_k}.
\label{eq:matrixelement}
\end{eqnarray}
The amplitude squared is composed of the propagators $\Delta(j)$, Eq.~(\ref{eq:propagators}),
of particle $j = t$, $\tilde\chi^\pm_i$, or $W$, 
and the un-normalized spin density matrices
$\rho_{D  }(\tilde b )$, $\rho_{D_1}(\tilde\chi^\pm_i)$, 
$ \rho_{D_2}(t)$, and $\rho_{D_3}(W)$,
with the helicity indices $\lambda_i,\lambda^\prime_i$ 
of the chargino, the helicity indices 
$\lambda_t,\lambda^\prime_t$ of the top quark, and those
of the $W$ boson, $\lambda_k,\lambda^\prime_{k}$.

The density matrices can be expanded in terms of the Pauli matrices
\begin{eqnarray} 
\rho_{D  }(\tilde b)_{\lambda_i\lambda^\prime_i}^{\lambda_t\lambda^\prime_t}
 &=&
       \delta^{\lambda_t\lambda^\prime_t}~\delta_{\lambda_i\lambda^\prime_i}~D
      +\delta_{\lambda_i\lambda^\prime_i}~
       (\sigma^a )^{\lambda_t\lambda^\prime_t}~\Sigma_D^a
      +\delta^{\lambda_t\lambda^\prime_t}~
       (\sigma^b)_{\lambda_i\lambda^\prime_i}~\Sigma_D^b
        \nonumber \\[2mm]
     && 
      +(\sigma^a)^{\lambda_t\lambda^\prime_t}~
       (\sigma^b)_{\lambda_i\lambda^\prime_i}~
       \Sigma_D^{ab},
\label{eq:rhoD}\\[2mm]
\rho_{D_1}(\tilde\chi^\pm_i)^{\lambda^\prime_i\lambda_i}&=&
          \delta^{\lambda^\prime_i\lambda_i}~D_1+
        (\sigma^b)^{\lambda^\prime_i\lambda_i}~\Sigma^b_{D_1}, 
\label{eq:rhoD1}\\[2mm]
\rho_{D_2}(t)_{\lambda^\prime_t\lambda_t}^{\lambda^\prime_k\lambda_k} &=&
        \left[\delta_{\lambda^\prime_t\lambda_t}~D_2^{\mu\nu}+
            (\sigma^a)_{\lambda^\prime_t\lambda_t}~
                \Sigma^{a\,\mu\nu}_{D_2}\right] \varepsilon^{\lambda_k\ast}_{\mu}
                \varepsilon^{\lambda^\prime_k}_{\nu},
\label{eq:rhoD2}\\[2mm]
\rho_{D_3}(W)^{\lambda^\prime_k\lambda_k}&=&
        D_3^{\rho\sigma}~\varepsilon_{\rho}^{\lambda_k}~
          \varepsilon^{\lambda^\prime_k\ast}_{\sigma},
\label{eq:rhoD3}
\end{eqnarray}
with an implicit sum over $a,b=1,2,3$.

The polarization vectors 
$\varepsilon^{\lambda_k}_{\mu}$ of the $W$ boson
fulfil the completeness relation
\begin{eqnarray}
\sum_{\lambda_k} \varepsilon^{\lambda_k\ast}_{\mu}
\varepsilon^{\lambda_k}_{\nu}= -g_{\mu\nu}+
   \frac{p_{W, \mu}~p_{W, \nu}}{m_W^2},
\label{eq:Wcompleteness}
\end{eqnarray}
with $p^\mu_W\:\varepsilon^{\lambda_k}_{\mu}=0$.
Similarly the  spin four-vectors $s^a_{t},$ $a=1,2,3,$ for the top quark $t$,
and $s^b_{\chi_i^\pm},$ $b=1,2,3,$  for the chargino $\tilde\chi^\pm_i$, 
also fulfil completeness relations
\begin{eqnarray}
\sum_{a}s_t^{a,\,\mu} s_t^{a,\,\nu} = -g^{\mu\nu}
      + \frac{p_t^\mu p_t^\nu}{m_t^2},
\qquad
\sum_{b}s_{\chi_i^\pm}^{b,\,\mu} s_{\chi_i^\pm}^{b,\,\nu} = -g^{\mu\nu}
      + \frac{p_{\chi_i^\pm}^\mu p_{\chi_i^\pm}^\nu}{m_{\chi_i^\pm}^2},
\label{eq:completeness}
\end{eqnarray}
and they form an orthonormal set
\begin{eqnarray}
&&s^a_t\cdot s^b_t=-\delta^{ab}, \quad
s^a_t\cdot \hat p_t=0, \qquad
s^a_{\chi_i^\pm}\cdot s^b_{\chi_i^\pm}=-\delta^{ab}, \quad
s^a_{\chi_i^\pm}\cdot \hat p_{\chi_i^\pm}=0,
\end{eqnarray}
with the notation
$\hat p^{\mu} = p^{\mu}/m$.

The expansion coefficients of the 
matrices, Eqs.~(\ref{eq:rhoD})-(\ref{eq:rhoD2}), are 
%
%
%
%
\begin{eqnarray} 
D &=& 
      \frac{g^2}{2}\left(|l^{\tilde b}_{mi}|^2+|k^{\tilde b}_{mi}|^2\right) 
      (p_t\cdot p_{\chi_i^\pm} )
      -g^2 {\rm Re}\{ l^{\tilde b}_{mi} (k^{\tilde b}_{mi})^\ast \}m_t m_{\chi_i^\pm},
\label{eq:D} \\ [2mm]
\Sigma^a_{D} &=& \,^{\;\,-}_{(+)}
        \frac{g^2}{2}\left(|l^{\tilde b}_{mi}|^2-|k^{\tilde b}_{mi}|^2\right)
        m_t(p_{\chi_i^\pm}\cdot s^a_{t}),
\label{eq:SigmaaD} \\ [2mm]
\Sigma^b_{D} &=& \,^{\;\,-}_{(+)}
        \frac{g^2}{2}\left(|l^{\tilde b}_{mi}|^2-|k^{\tilde b}_{mi}|^2\right)
        m_{\chi_i^\pm}(p_t\cdot s^b_{{\chi_i^\pm}}),
\label{eq:SigmabD} \\ [2mm]
\Sigma^{ab}_{D} &=& 
         \frac{g^2}{2}\left(|l^{\tilde b}_{mi}|^2+|k^{\tilde b}_{mi}|^2\right)
         (s^a_{t}\cdot s^b_{{\chi_i^\pm}})m_t m_{\chi_i^\pm}
        \nonumber \\ [2mm] &&
         + g^2{\rm Re}\{ l^{\tilde b}_{mi} (k^{\tilde b}_{mi})^\ast \}
\left[  (s^a_{t}\cdot   p_{\chi_i^\pm}) (s^b_{\chi_i^\pm}\cdot p_t)
       -(s^a_{t}\cdot s^b_{\chi_i^\pm}) (p_{\chi_i^\pm}  \cdot p_t)
\right]
        \nonumber \\ [2mm] &&
        -g^2 {\rm Im}\{ l^{\tilde b}_{mi} (k^{\tilde b}_{mi})^\ast \}
         [s^{a}_{t},~ p_t,~ s^{b}_{\chi_i^\pm},~p_{\chi_i^\pm}],
\label{eq:sigmaabD} \\ [2mm]
D_1 &=& \frac{g^2}{2} |V_{i1}|^2 (m_{\chi_i^\pm}^2 -m_{\tilde\nu_\ell}^2 ),
\label{eq:D1} \\ [2mm]
\Sigma^b_{D_1} &=& \,^{\;\,+}_{(-)} g^2 |V_{i1}|^2 
m_{\chi_i^\pm} (s^b_{\chi_i^\pm} \cdot p_{\ell_1}),
\label{eq:SigmabD1} \\ [2mm]
{D_2}^{\mu\nu}&=& 
   \frac{g^2}{2}\left[
      p^{\mu}_b p^{\nu}_t +   p^{\nu}_b p^{\mu}_t
        -( p_b \cdot p_t) g^{\mu\nu}
               \right]
        \,^{\;\,+}_{(-)}\frac{g^2}{2}i
     \varepsilon^{\mu\alpha\nu\beta}p_{t, \, \alpha}~p_{b,\,\beta},
\label{eq:D2} \\ [2mm]
\Sigma^{a\, \mu\nu}_{D_2} &=& 
   \,^{\;\,-}_{(+)}
    \frac{g^2}{2}m_t \left\{ \left[
      p^{\mu}_b s^{a,\, \nu}_t +  p^{\nu}_b s^{a,\,\mu}_t
        -( p_b \cdot s^a_t) g^{\mu\nu}
               \right]
        \,^{\;\,+}_{(-)}i
     \varepsilon^{\mu\alpha\nu\beta}s_{t, \, \alpha}^a~p_{b,\,\beta}
                \right\},\quad
\label{eq:SigmaD2}\\ [2mm]
D_3^{\rho\sigma}&=& 
     g^2\left[p^{\rho}_{\ell_2} p^{\sigma}_{\nu_\ell} 
             +p^{\sigma}_{\ell_2} p^{\rho}_{\nu_\ell}
            -( p_{\ell_2} \cdot p_{\nu_\ell}) g^{\rho\sigma}
         \right]
         \,^{\;\,+}_{(-)} g^2
     i \varepsilon^{\rho\alpha\sigma\beta}p_{\ell_2,\,\alpha}~p_{\nu_\ell,\,\beta},
\label{eq:D4} 
\end{eqnarray}
with the couplings as defined in Appendix~\ref{Lagrangians and couplings},
and the short hand notation
\begin{equation}
     [p_{1},p_2,p_{3},p_{4}]	\equiv 
	\varepsilon_{\mu\nu\alpha\beta}~
	p_{1}^{\mu}~p_2^{\nu}~p_{3}^{\alpha}~p_{4}^{\beta};
      \quad  \varepsilon_{0123}=1.
\label{eq:epsilonB}      
\end{equation}
The coefficients for sbottom decay, Eqs.~(\ref{eq:D})-(\ref{eq:sigmaabD}),
are obtained from those of stop decay given in Ref.~\cite{Deppisch:2009nj},
by the replacements of the couplings
$a^{\tilde t}_{mi}\to l^{\tilde b}_{mi}$, and 
$b^{\tilde t}_{mi}\to k^{\tilde b}_{mi}$.
The signs in parentheses hold for the charge conjugated processes, that is
$\tilde b_m^\ast \to \bar t \tilde\chi^+_i$
in Eqs.~(\ref{eq:SigmaaD}) and (\ref{eq:SigmabD}),
$\tilde\chi^+_i \to \ell_1^+ \tilde\nu_\ell^\ast$ in Eq.~(\ref{eq:SigmabD1}),
$\bar t \to\bar b  W^-$ in Eqs.~(\ref{eq:D2}) and (\ref{eq:SigmaD2}),
and finally $W^- \to \bar\nu_\ell\, \ell_2^-$ in Eq.~(\ref{eq:D4}).

Inserting the density matrices, Eqs.~(\ref{eq:rhoD})-(\ref{eq:rhoD3})
into Eq.~(\ref{eq:matrixelement}), we obtain 
\begin{eqnarray}
    |T|^2=4~|\Delta(t)|^2~|\Delta(\tilde\chi^\pm_i)|^2~|\Delta(W)|^2
 \times \nonumber \\ [3mm]
\left[
 D~D_1~D_2^{\rho\sigma}
+ D_1~\Sigma^a_{D}~\Sigma^{a \,\rho\sigma}_{D_2}
+ \Sigma^b_{D}~\Sigma^{b}_{D_1}~D_2^{\rho\sigma}
+  \Sigma^{ab}_{D}~\Sigma^{b}_{D_1}~\Sigma_{D_2}^{a\,\rho\sigma}
\right]
D_{3\,\rho\sigma},
\label{eq:matrixelement2}
\end{eqnarray}
with an implicit sum over $a,b$.
The amplitude squared $|T|^2$ is now composed into an unpolarized part (first summand), 
into the spin correlations of the top (second summand), those of the chargino 
(third summand), and into the spin-spin correlations of top and chargino
(fourth summand), in Eq.~(\ref{eq:matrixelement2}).

With the completeness relation for the $W$ polarization
vectors, Eq.~(\ref{eq:Wcompleteness}), we find
\begin{eqnarray}
D_2^{\rho\sigma}~D_{3\,\rho\sigma}&=& 
2g^4(p_t\cdot p_{\ell_2})(p_b\cdot p_{\nu_\ell}),
\label{eq:d2d3}\\[2mm]
\Sigma_{D2}^{a\,\rho\sigma}~D_{3\,\rho\sigma}&=& 
\,^{\;\,-}_{(+)} 2m_tg^4(s^a_t\cdot p_{\ell_2})(p_b\cdot p_{\nu_\ell}),
\label{eq:si2ad3}
\end{eqnarray}
and the sign in parenthesis for the charge conjugated decay, 
$\bar t \to\bar b  W^-$, $W^- \to \bar\nu_\ell \ell_2^-$.
By also using the completeness relations for the top and
chargino spin vectors, Eq.~(\ref{eq:completeness}),
the products in Eq.~(\ref{eq:matrixelement2}) can be written as
\begin{eqnarray}
 \Sigma^a_{D}~\Sigma^{a \,\rho\sigma}_{D_2}~D_{3\,\rho\sigma}
&=&
  g^6\left(|l^{\tilde b}_{mi}|^2 - |k^{\tilde b}_{mi}|^2\right)
   (p_b\cdot p_{\nu_\ell})  \nonumber\\ 
   &&\times\left[
      (p_{\chi_i^\pm}\cdot p_t)(p_{\ell_2}\cdot p_t)
     -m_t^2(p_{\chi_i^\pm}\cdot p_{\ell_2})
  \right],  \label{eq:product1}\\
\Sigma^b_{D}~\Sigma^{b}_{D_1} 
&=&  
  \frac{g^4}{2}\left(|l^{\tilde b}_{mi}|^2 - |k^{\tilde b}_{mi}|^2\right)
   |V_{i1}|^2 \nonumber\\ 
  &&\times\left[
      m_{\chi_i^\pm}^2(p_t\cdot p_{\ell_1})
      -(p_{\chi_i^\pm}\cdot p_t)(p_{\ell_1}\cdot p_{\chi_i^\pm})
  \right], \label{eq:product2}
\\
\Sigma^{ab}_{D}~\Sigma^{b}_{D_1}~\Sigma_{D_2}^{a\,\rho\sigma}~D_{3\,\rho\sigma}
&=&
g^8\left(|l^{\tilde b}_{mi}|^2 + |k^{\tilde b}_{mi}|^2\right)
|V_{i1}|^2
\left[
    (p_{\chi_i^\pm}\cdot p_{\ell_1})(p_{\chi_i^\pm}\cdot p_{\ell_2}) m_t^2
\right.
\nonumber \\ [2mm]
&&
   +(p_t\cdot p_{\ell_1})(p_t\cdot p_{\ell_2}) m_{\chi_i^\pm}^2
   -(p_{\ell_1}\cdot p_{\ell_2}) m_{\chi_i^\pm}^2 m_t^2
 \nonumber \\ [2mm]
&&
\left.
  -(p_{\chi_i^\pm}\cdot p_t)(p_t\cdot p_{\ell_2})(p_{\chi_i^\pm}\cdot p_{\ell_1})
  \right]
(p_b\cdot p_{\nu_\ell})
 \nonumber \\ [2mm]
&&
  + 2g^8{\rm Re}\{ l^{\tilde b}_{mi} (k^{\tilde b}_{mi})^\ast \}
  |V_{i1}|^2  m_{\chi_i^\pm} m_t(p_b\cdot p_{\nu_\ell})
\nonumber \\ [2mm]
&&
\times
\left[
     (p_{\chi_i^\pm}\cdot p_t)(p_{\ell_1}\cdot p_{\ell_2})
    -(p_{\chi_i^\pm}\cdot p_{\ell_2})(p_t\cdot p_{\ell_1})
 \right]
\nonumber \\ [2mm]
&&
   - 2g^8{\rm Im}\{ l^{\tilde b}_{mi} (k^{\tilde b}_{mi})^\ast \}
   |V_{i1}|^2 m_{\chi_i^\pm} m_t(p_b\cdot p_{\nu_\ell})
         [p_{\tilde b}, p_t, p_{\ell_1}, p_{\ell_2}],
\nonumber \\ 
\label{CPterm}
\end{eqnarray}
with the short hand notation Eq.~(\ref{eq:epsilonB}).
There is no sign change in the terms Eqs.~(\ref{eq:product1})-(\ref{CPterm}) 
for the charge conjugated process $\tilde b_m^\ast \to \bar t \tilde\chi^+_i$,
with the subsequent decays $\tilde\chi^+_i \to \ell_1^+ \tilde\nu _\ell^\ast$, and
$\bar t \to\bar b  W^-$, $W^- \to \bar\nu_\ell \ell_2^-$.

%
\section{Sbottom decay widths and asymmetry}
  \label{Sbottom decay width}

The partial decay width for the sbottom decay $\tilde b_m \to t \tilde\chi_i^\pm$
is~\cite{Bartl:2003he}
\begin{equation}
\Gamma(\tilde b_m \to t \tilde\chi_i^\pm)=
\frac{\sqrt{\lambda (m^2_{\tilde b},m^2_t,m^2_{\chi^\pm_i})}\; }
{4 \pi m^3_{\tilde b}}D,
\label{eq:widtht}
\end{equation}
with the decay function $D$ given in Eqs.~(\ref{eq:D}).
For the decay $\tilde b_m \to b \tilde\chi_j^0$ we have~\cite{Bartl:2003he}
\begin{equation}
\Gamma(\tilde b_m \to b \tilde\chi_j^0)=
\frac{ (m^2_{\tilde b} -m^2_{\chi^0_j})^2 }
{16 \pi m^3_{\tilde b}}g^2
\left(|a^{\tilde b}_{mj}|^2 + |b^{\tilde b}_{mj}|^2\right),
\label{eq:widthb}
\end{equation}
whith the approximation $m_b= 0$. The
 sbottom-bottom-neutralino couplings are~\cite{Bartl:2003he}
\begin{eqnarray}
a^{\tilde b}_{mj} &=&
 {\mathcal R}^{\tilde b \ast}_{m1}\, f^L_{bj}
+{\mathcal R}^{\tilde b \ast}_{m2}\, h^R_{bj}, \qquad
b^{\tilde b}_{mj} \:=\:
 {\mathcal R}^{\tilde b \ast}_{m1}\, h^L_{bj}
+{\mathcal R}^{\tilde b \ast}_{m2}\, f^R_{bj},\\[2mm]
 f^L_{bj}&=&
 -\sqrt{2}\bigg[\frac{1}{\cos
        \theta_w}\left(-\frac{1}{2}+\frac{1}{3}\sin^2\theta_w\right)N_{j2}-
        \frac{1}{3}\sin \theta_w N_{j1}\bigg],
\label{eq:fld}\\[2mm]
f^R_{bj}&=&
        \frac{\sqrt{2}}{3} \sin \theta_w
        \left(\tan\theta_w N_{j2}^\ast-N_{j1}^\ast\right),\\[2mm]
h^L_{bj} &=& (h^R_{bj})^\ast \;=\;
-Y_b(\cos\beta N_{j3}^\ast + \sin\beta N_{j4}^\ast ),
\label{eq:sbneutcouplings}
\end{eqnarray}
with the sbottom diagonalization matrix ${\mathcal R}^{\tilde b}$,  
Eq.~(\ref{eq:rsbottom}),
and $N$ the diagonalization matrix for the neutralino matrix 
in the photino, zino, higgsino basis, see Ref.~\cite{Deppisch:2009nj}

The sbottom decay width for the complete decay chain,
Eqs.~(\ref{eq:decaySbottom})-(\ref{eq:decayTop}), is  given by 
\begin{equation}
\Gamma(\tilde b \to \nu_\ell\,\tilde\nu_\ell\,\tilde\chi_i^\pm\, b\,\ell_1\ell_2)=
        \frac{1}{2 m_{\tilde b}}\int|T|^2
      d{\rm Lips}(s_{\tilde b}\,;
       p_{\nu_\ell},p_{\tilde\nu_\ell},p_{\chi_i^\pm},p_b,
       p_{\ell_1},p_{\ell_2}),
\label{eq:width}
\end{equation}
with the phase-space element $d{\rm Lips}$ given in 
Eq.~(\ref{eq:phasespace}).

We obtain an explicit expression for the asymmetry,
if we insert the amplitude squared $|T|^2$, Eq.~(\ref{eq:matrixelement2}),
into Eq.~(\ref{eq:Toddasym}),
\begin{equation}
{\mathcal A}=
  \frac{\int {\rm Sign}({\mathcal E})\,
 \Sigma^{ab}_{D}~\Sigma^{b}_{D_1}~\Sigma_{D_2}^{a\,\rho\sigma}
~D_{3\,\rho\sigma}~
  d{\rm Lips}  }
{\int D~D_1~D_2^{\rho\sigma}~D_{3\,\rho\sigma}~
d{\rm Lips} },
\label{eq:Adependence}
\end{equation}
where we have already used the narrow width approximation of the propagators,
see Eq.~(\ref{narrowwidth}).
In the numerator, only the spin-spin terms of the
amplitude squared remain, since only they contain 
the epsilon product ${\mathcal E}$, see Eq.~(\ref{eq:epsilon}).
The other terms vanish due to the phase-space integration
over the sign of the epsilon product, ${\rm Sign}({\mathcal E})$.
In the denominator, all spin and spin-spin correlation
terms vanish, and only the spin-independent parts contribute.
Inserting now the explicit expressions of the terms
Eqs.~(\ref{eq:D})-(\ref{CPterm})
into the formula for the asymmetry, Eq.~(\ref{eq:Adependence}), we find 
\begin{equation}
{\mathcal A}=  \eta \;
	\frac{\int {\rm Sign}({\mathcal E})
	(p_b\cdot p_{\nu_\ell})\,
	[p_{\tilde b},p_t,p_{\ell_1},p_{\ell_2}]~
	d{\rm Lips}  }{(p_{\chi_i^\pm}\cdot p_{\ell_1})\int
	(p_t\cdot p_{\ell_2})
	(p_b\cdot p_{\nu_\ell})~
	d{\rm Lips} },
\label{eq:Adependence3}
\end{equation}
with the coupling function
\begin{equation}
\eta = \frac{{\rm Im}\{ l^{\tilde b}_{mi} (k^{\tilde b}_{mi})^\ast \}}
      { \frac{1}{2}\left(|l^{\tilde b}_{mi}|^2 + |k^{\tilde b}_{mi}|^2\right)
     \frac{m_{\tilde b}^2 - m_{\chi_i^\pm}^2 - m_t^2}
    {2 m_{\chi_i^\pm} m_t}
     - {\rm Re}\{ l^{\tilde b}_{mi} (k^{\tilde b}_{mi})^\ast \}}.
\label{eq:couplingfunct2}
\end{equation}

\section{Theoretical statistical significance}
\label{sec:Significances}

%
Assuming that the fluctuations of the signal rate are binomially distributed with 
the selection probability $p=1/2(\mathcal{A}+1)$, 
the significance of an asymmetry is~\cite{Deppisch:2009nj}
\begin{eqnarray}
	{\mathcal S} = 
	\frac{|{\mathcal A}|}{\sqrt{1-{\mathcal A}^2}} 
	\sqrt{\sigma{\mathcal L}},
\label{eq:significance}
\end{eqnarray} 
with the integrated LHC luminosity ${\mathcal L}$, and the 
cross section $\sigma$ of sbottom production and decay as defined below.
The statistical significance  $\mathcal{S}$ is equal to the number of standard 
deviations that the asymmetry can be statistically determined to be non-zero. 
For example a value of $\mathcal{S} = 1$ implies a measurement at the $68\%$ confidence level. 
The minimal required luminosity is then
\begin{eqnarray}
{\mathcal L} &=& \frac{1}{\sigma}\left( \frac{1}{{\mathcal A^2}} -1 \right).
\label{eq:lumi2}
\end{eqnarray}
The cross section for sbottom production and decay  is 
\begin{eqnarray}
	\sigma &=& F_N \times
	\sigma(pp\to\tilde b_m\tilde b_m^\ast)  
	\times {\rm Br}(\tilde b_m \to t  \tilde\chi^-_i) 
	\times {\rm Br}(t\to b  W) \times {\rm Br}(W \to \nu_e e)
 \nonumber\\[2mm]
	&& \phantom{F_N }
        \times {\rm Br}(\tilde\chi^-_i\to  e^+\tilde \nu_e ) 
	\times {\rm Br}(\tilde \nu_e \to \tilde\chi^0_1 \nu_e),
	\label{eq:crosssection}
\end{eqnarray}
for $m=1,2$, and $i=1,2$. For $m,i$ fixed, the combinatorial factor $F_N$ takes into 
account the possible $W$ and  chargino $\tilde\chi^-_i$  decays into leptons with 
different flavors.
We assume that the branching ratios do not depend on the flavor, i.e.,
${\rm Br}(W \to \nu_e e) = {\rm Br}(W \to \nu_\mu \mu)$,
and $
 {\rm Br}(\tilde\chi^-_i\to e^{-}\tilde \nu_e)=
 {\rm Br}(\tilde\chi^-_i\to \mu^{-}\tilde \nu_\mu)$.
The   factor is thus $F_N=4$, if we sum  the lepton flavors $e, \mu$.
We further assume $ {\rm Br}(\tilde \nu_e \to \tilde\chi^0_1 \nu_e)=1$,
which applies to our SUSY scenarios considered.

\medskip
For the calculation of the sbottom decay widths and branching ratios, we use the
formulas as given in Eqs.~(\ref{eq:widtht}), (\ref{eq:widthb}).
For the calculation of the chargino decay widths and branching ratios, we  
consider the two-body decays~\cite{Kittel:2004rp}  
\begin{eqnarray}  
{\tilde\chi}_1^\pm&\to&  
    \ell\tilde\nu_\ell,\;   
            \tau\tilde\nu_\tau, \;
           \tilde\ell_L    \nu_\ell, \;
           \tilde\tau_{1,2}\nu_\tau, \;
              W \tilde\chi_1^0,
\quad \ell = e,\mu.
\label{chardecyas} 
\end{eqnarray}  
We neglect three-body decays, which are suppressed by phase space.

\end{appendix}



\end{document}